\documentclass[conference]{IEEEtran}
\IEEEoverridecommandlockouts

\usepackage{cite}
\usepackage{amsmath,amssymb,amsfonts}
\usepackage{algorithmic}
\usepackage{graphicx}
\usepackage{multirow}
\usepackage{booktabs}
\usepackage{subfigure}
\usepackage{textcomp}
\usepackage{xcolor}
\usepackage[table,xcdraw]{xcolor}
\usepackage[most]{tcolorbox}
\usepackage[normalem]{ulem}
\useunder{\uline}{\ul}{}
\usepackage{url}

\def\BibTeX{{\rm B\kern-.05em{\sc i\kern-.025em b}\kern-.08em
    T\kern-.1667em\lower.7ex\hbox{E}\kern-.125emX}}

\newcommand\black[1]{\textcolor{black}{#1}}

\begin{document}

\setlength{\textfloatsep}{7pt plus 2pt minus 2pt} 

\title{Assessing Superposition-Targeted Coverage Criteria for Quantum Neural Networks}

\author{\IEEEauthorblockN{Minqi Shao}
\IEEEauthorblockA{
\textit{Kyushu University}\\
Fukuoka, Japan \\
shao.minqi.229@s.kyushu-u.ac.jp}
\and
\IEEEauthorblockN{Jianjun Zhao}
\IEEEauthorblockA{
\textit{Kyushu University}\\
Fukuoka, Japan \\
zhao@ait.kyushu-u.ac.jp}
}

\maketitle

\begin{abstract}
Quantum Neural Networks (QNNs) have achieved initial success in various tasks by integrating quantum computing and neural networks. However, growing concerns about their reliability and robustness highlight the need for systematic testing. Unfortunately, current testing methods for QNNs remain underdeveloped, with limited practical utility and insufficient empirical evaluation.
As an initial effort, we design a set of superposition-targeted coverage criteria to evaluate QNN state exploration embedded in test suites.
To characterize the effectiveness, scalability, and robustness of the criteria, we conduct a comprehensive empirical study using benchmark datasets and QNN architectures. We first evaluate their sensitivity to input diversity under multiple data settings, and analyze their correlation with the number of injected faults. We then assess their scalability to increasing circuit scales. The robustness is further studied under practical quantum constraints including insufficient measurement and quantum noise. The results demonstrate the effectiveness of quantifying test adequacy and the potential applicability to larger-scale circuits and realistic quantum execution, while also revealing some limitations. Finally, we provide insights and recommendations for future QNN testing.
\end{abstract}



\section{Introduction}

Quantum computing \cite{de2024quantum} and machine learning \cite{ml_survey} have advanced rapidly in recent years, giving rise to a frontier field, Quantum Machine Learning (QML)~\cite{qml}. Among QML approaches, Quantum Neural Networks (QNNs) have been applied to tasks such as image classification~\cite{qnn_survey} and sequential data learning~\cite{quantumnlp}, offering quantum potentials in time-complexity speedup~\cite{timecomplexity} and model compression~\cite{compression}. Both theoretical and empirical research on QNNs are currently active, with major efforts focusing on architecture design \cite{qcnn}, property analysis \cite{qleet}, and optimization problems \cite{bp}.

Despite these promises and advances, recent studies have shown that QNNs are highly vulnerable to adversarial perturbations~\cite{qaml} and backdoor attacks~\cite{qdoor}, raising concerns about their practical reliability and safety in future applications. Their robustness is further threatened by quantum-unique characteristics such as system dimension~\cite{quantumvulnerability} and quantum noise~\cite{noiseforrobust}. Unfortunately, current QNN testing remains largely underexplored. The only dedicated testing framework, QuanTest~\cite{quantest}, uses entanglement as a guide to generate high-quality adversarial samples. However, the core element, MW entanglement~\cite{mw}, is practically infeasible due to quantum state collapse, and its estimation incurs additional variance and resource overhead~\cite{haug2023scalable}. Insufficient evaluation under realistic quantum execution, including limited measurement shots and quantum noise, limits its practical value for QNN testing. Moreover, interpretability requires further improvement, especially in defining entangled neurons.

In classical software testing, coverage criteria \cite{traditionalcoverage} provide a practical surrogate for test adequacy by quantifying how thoroughly a test suite exercises program elements such as statements and branches, and are widely used to guide test generation or improve existing test suites. In the field of testing quantum programs (QPs) and deep neural networks (DNNs), coverage criteria have also received increasing attention in recent years. For QPs, tools such as Quito~\cite{quito} and gate branch coverage~\cite{gatecoverage} rely on explicit program specifications or branch enumeration, which requires manual effort and pre-defined program logic. These assumptions are difficult to satisfy for learning-based QNNs, where input--output spaces are much broader and test oracles are often lacking. This gap motivates automated criteria that are sensitive to the diversity of model behaviors. For DNNs, neuron-based criteria~\cite{deepxplore,deepgauge,surprise,npc,Liu2023NeuronAC} explore individual or collective activation patterns within models. However, these notions are inapplicable to QNNs constructed from quantum gates rather than neurons. More importantly, state collapse caused by mid-circuit measurement and irreversible quantum operations renders many white-box criteria infeasible in practice.
These limitations highlight the need for new testing criteria tailored to QNNs at a finer granularity and black-box level.

When evaluating testing techniques, recent research has looked beyond effectiveness and increasingly examined how quantum-specific factors affect testing performance in practice \cite{qst,qst2}. One major concern is scalability, which focuses on whether testing remains applicable as circuit complexity grows. Circuit complexity directly impacts execution efficiency and computational cost, and is commonly characterized by the number of qubits, circuit depth, and gate count \cite{20emp}. Another concern is robustness, which examines whether testing remains stable under quantum-specific stochasticity induced by measurement and noise. For measurement, the number of shots determines the reliability of circuit outputs and thus affects testing results. However, the community still lacks a common reference for this setting, and typically adopts a fixed budget or a statistically minimum bound \cite{Ye2025IsME}. In the Noisy Intermediate-Scale Quantum (NISQ) era, inherent noise of quantum devices further introduces perturbations that threaten stable testing. As QNNs can be regarded as a specialized form of quantum programs, evaluations of QNN-oriented testing techniques should incorporate these factors to provide a more comprehensive assessment.

To make an initial effort towards QNN testing, we propose a set of superposition-targeted coverage criteria for QNNs. It includes three multi-granularity metrics, \emph{$K$-cell State Coverage (KSC)}, \emph{State Corner Coverage (SCC)}, and \emph{Top-$k$ State Coverage (TSC)}, to evaluate the sufficiency of state exploration within the output space of QNNs. KSC and SCC quantify the major and corner model behaviors covered by given test suites, while TSC sheds light on the impact of individual states on model decision logic. These criteria are based on superposition-derived probabilistic statistics, which are directly measurable and expressive in capturing behavioral changes. This design avoids the inaccuracy and resource overhead associated with more intractable quantum properties. To assess the effectiveness, scalability, and robustness of the proposed criteria, we conduct a comprehensive empirical study on benchmark datasets and representative QNN architectures. We construct two groups of test suites from different perspectives, such as out-of-distribution data. The results demonstrate sensitivity to different suites and scalability to larger-scale circuits. Moreover, given a confidence level and an error bound, the criteria maintain stable under finite measurement shots and simulated noise. We further identify several underlying limitations, and summarize insights and recommendations for future research on QNN testing and its evaluation.

This paper makes the following contributions:
\begin{itemize}
    \item We propose a set of superposition-targeted coverage criteria for QNNs at different granularities to quantify test adequacy in the output space.
    \item We conduct a comprehensive empirical study on two benchmark datasets and four representative QNN architectures to assess the effectiveness, scalability, and robustness of coverage criteria, with particular attention to quantum-specific factors.
    \item We provide guidance for future research on developing testing criteria for QNNs and offer recommendations for conducting evaluation in the quantum context.
\end{itemize}

\section{Background and related work}

\subsection{Basic Knowledge of Quantum Computing}

\noindent 
\textbf{Qubit and Quantum State.} A quantum bit (\textit{qubit}) is the basic unit of quantum computing, analogous to classical bits. Unlike classical bits, qubits can exist in a \textit{superposition} of computational basis states as a linear combination. A quantum state $|\phi\rangle$ of a qubit is typically represented as 
$|\phi\rangle = \alpha|0\rangle + \beta|1\rangle$, where $\alpha, \beta \in \mathbb{C}$ are \textit{amplitudes} that satisfy $|\alpha|^2+|\beta|^2=1$. A system of $n$ qubits exists in a superposition of $2^n$ basis states, corresponding to the binary strings from $0$ to $2^n-1$.

\begin{figure}[t]
  \centering
  \includegraphics[width=0.75\linewidth]{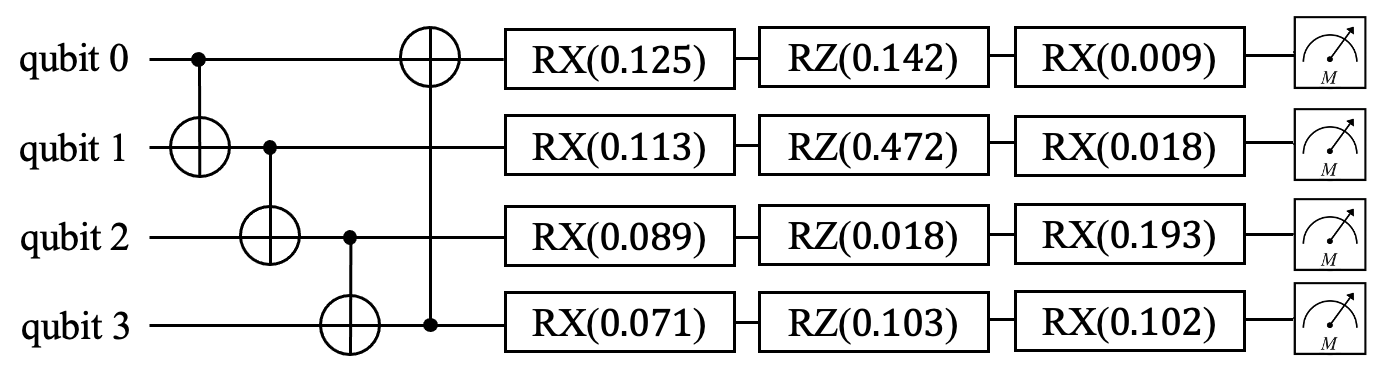}
    \caption{Example of a parameterized quantum circuit (PQC).}
    \label{fig:pqc}
\end{figure}

\noindent \textbf{Quantum Gates and Circuits.} Quantum gates are unitary operations that manipulate qubits by modifying amplitudes or creating entanglement. A quantum gate represents a linear, reversible, and norm-preserving transformation on the corresponding quantum state. For example, single-qubit rotation gates including $R_x, R_y$, and $R_z$ apply angle-parameterized rotations, while the Hadamard gate generates superpositions and the CNOT gate introduces entanglement between qubits.
A quantum circuit consists of a sequence of quantum gates and typically ends with measurements. Figure~\ref{fig:pqc} illustrates a simple circuit composed of rotation and CNOT gates.

\noindent
\textbf{Measurement.}
In practice, direct access to quantum state amplitudes is strictly limited. Quantum measurement extracts classical information from a quantum system by collapsing a superposition into a definite state based on the corresponding amplitude, in an irreversible manner. This process introduces inherent stochasticity that distinguishes quantum computation from classical execution.

\noindent
\textbf{Entanglement.}
Entanglement is a fundamental quantum property without classical counterparts, where the states of two or more qubits become correlated regardless of physical distance. When qubits are entangled, measuring one qubit constrains the state of the others. This property plays a crucial role in quantum algorithms achieving quantum advantage.

\subsection{Quantum Neural Networks}
\label{sec:qnn}

\begin{figure}[t]
  \centering
  \includegraphics[width=1.\linewidth]{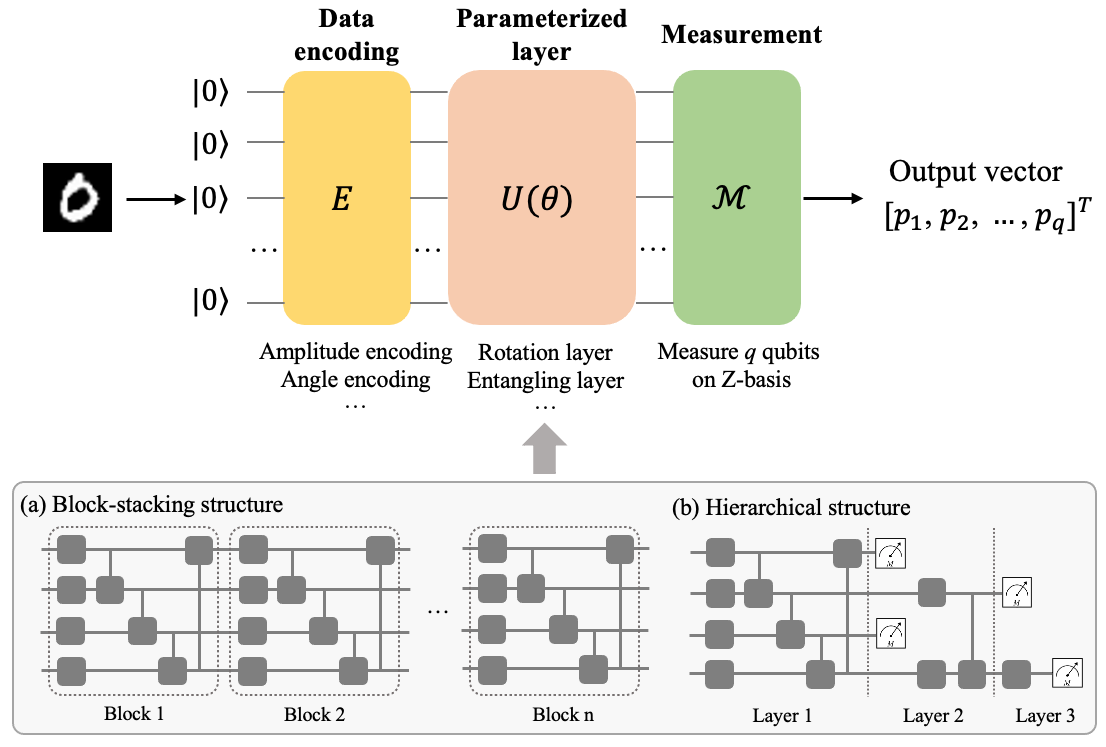}
    \caption{A general structure of QNN.}
    \label{fig:qnn}
\end{figure}

Quantum neural networks (QNNs)~\cite{qnn_survey} are typically constructed from parameterized quantum circuits (PQCs) with predefined structures and tunable parameters optimized for specific tasks. The example in Figure~\ref{fig:pqc} illustrates a linear entangling structure of CNOT gates and a rotation layer of parameterized rotation gates.
As shown in Figure~\ref{fig:qnn}, a typical QNN consists of three components: a data encoding layer to encode classical data, a parameterized circuit layer to transform quantum states, and a measurement layer to extract classical outputs.
In the data encoding layer, \textit{amplitude encoding} maps classical features to quantum state amplitudes, requiring fewer qubits but deeper circuits, while \textit{angle encoding} encodes features as rotation parameters, enabling efficient implementation at the cost of more qubits. For the parameterized layer, two representative designs are the \textit{block-stacking structure}~\cite{qcl,drnn} which repeatedly stacks the same block, and the \textit{hierarchical structure}~\cite{qcnn,hcqc} which reduces the freedom of qubits as circuit deepens. In practice, these circuits are typically organized as alternating rotation and entangling layers. Finally, given a finite number of shots, measurement collapses the quantum state, and expectation values on selected qubits are used as QNN outputs~\cite{hcqc}.

Despite their structural differences from DNNs, current QNN training pipelines remain largely classical. Loss computation and parameter updates are performed classically, while quantum devices or simulators are only used for forward execution to obtain model outputs.

\subsection{Coverage Criteria for Quantum Programs and Deep Neural Networks}

Quantum Programs (QPs) are written in quantum programming languages and typically contain both classical and quantum subroutines. To address challenges produced by probabilistic measurement, several coverage criteria have been proposed. Quito~\cite{quito} defines coverage over program inputs and outputs together with two statistical test oracles. Gate Branch Coverage (GBC)~\cite{gatecoverage} measures exercised branches of controlled gates during execution. These criteria aim to enhance test adequacy and fault detection.
However, such criteria are not suitable for learning-based QNNs. They rely on program specifications with limited predefined input-output pairs and explicit branch structures, whose assumptions do not hold for QNNs with broad state spaces and learned decision logic.

In parallel, the reliability of deep neural networks (DNNs) has been widely studied, including robustness, correctness, and generalization~\cite{fgsm,generalization}. 
A variety of coverage criteria have been proposed to characterize model behavior exploration. DeepXplore~\cite{deepxplore} introduced neuron coverage, and DeepGauge~\cite{deepgauge} extended it with multi-granularity metrics. Subsequent work explored neuron combinations~\cite{deepct}, decision paths~\cite{npc}, and feature-level behaviors~\cite{deepfeature}. While these criteria provide insights into model behavior, their effectiveness in fault detection remains under active investigation~\cite{zhou2025evaluating}.
Nevertheless, these criteria are not directly applicable to QNNs, as their basic computational units are quantum gates rather than neurons. Moreover, due to superposition and entanglement, the contribution of individual gates to global behavior is difficult to interpret, and there is no commonly accepted notion of quantum neurons.

From the perspective of model accessibility, most existing coverage criteria are white-box and require full access to internal structures. This limitation is more severe in the quantum context, where mid-circuit measurements cause irreversible state collapse and incur additional resource costs, motivating the exploration of black-box testing approaches.


\section{Coverage Criteria for Quantum Neural Networks}

\begin{figure}[t]
    \centering
  \includegraphics[width=1.\linewidth]{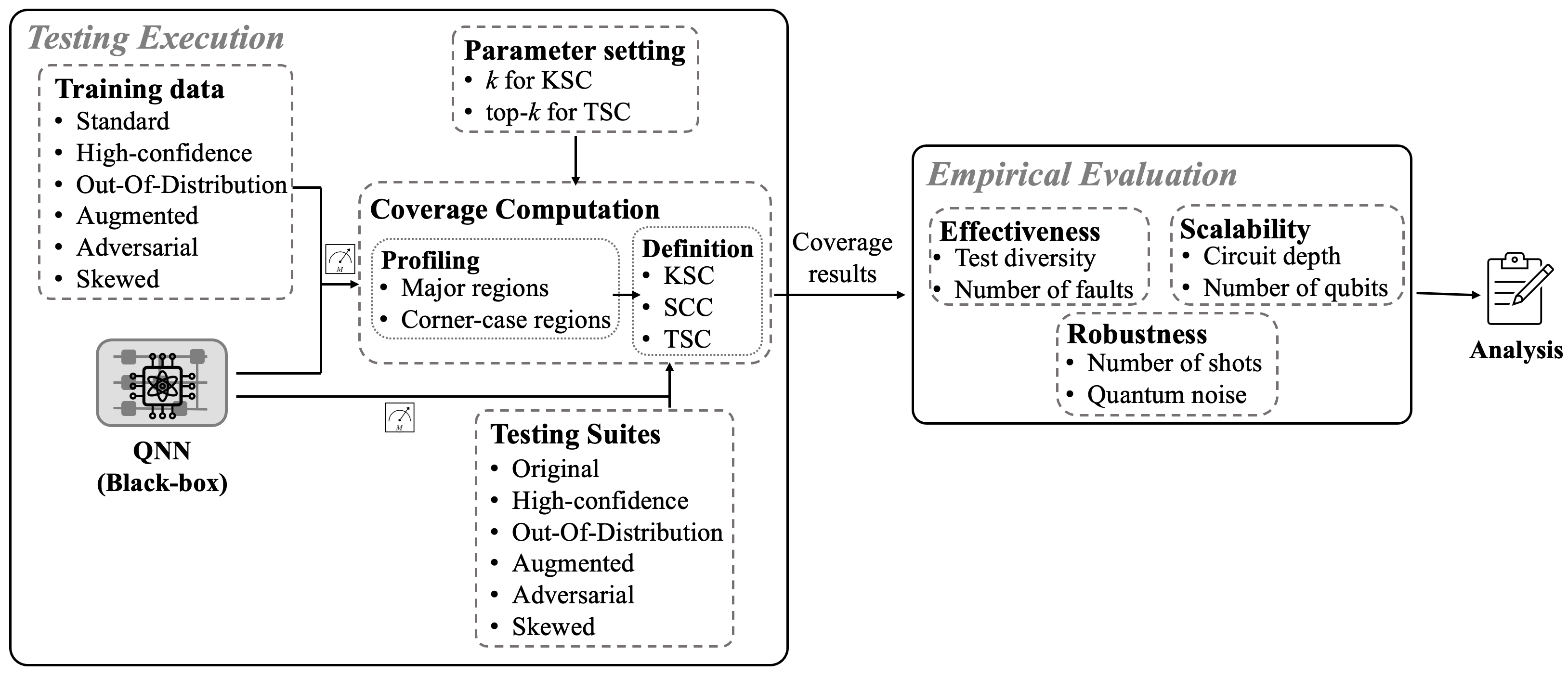}
    \caption{The overview of assessing coverage criteria for QNNs. During testing execution, it involves profiling model behaviors and computing coverage. Subsequently, empirical evaluation is conducted based on coverage results to analyze their effectiveness, scalability and robustness.}
    \label{fig:overview}
\end{figure}

The assessment workflow for coverage criteria is illustrated in \figurename \ref{fig:overview}. During testing execution, model behaviors are profiled using training data to characterize major regions, with corner-case regions defined as deviations beyond these regions. Coverage results are then computed for given test suites. The distribution of training and test sets may vary, including out-of-distribution or adversarial samples. Finally, coverage results are evaluated in terms of effectiveness, scalability and robustness.

\subsection{Design Principles}
For a $q$-qubit system, the quantum state lives in a $2^q$-dimensional Hilbert space. By the principle of superposition, it can be expressed as a linear combination of computational basis states. We design criteria based on this quantum property rather than original model outputs (e.g., logits or labels), to enable a more exploratory output space and non-trivial evaluation. Since superposition cannot be directly observed from a single measurement, we infer it from measurement statistics, namely the \textit{probability distribution} estimated from repeated measurements on a quantum state. It specifies the likelihood of each basis state being observed and provides an alternative to the underlying superposition. Compared with high-level quantities such as entanglement, this probabilistic information is \textit{the primary} statistic directly extractable from quantum systems. It does not require additional estimation procedures, making it a natural and feasible indicator of model behaviors. We operate in a \textbf{black-box} way, relying on the probability distributions of the output states. This is suitable for quantum-context tasks where mid-circuit measurements can cause the irreversible collapse of quantum states.

However, the number of basis states grows exponentially with the number of qubits, and the probability of each state theoretically ranges from 0 to 1. This vast and continuous solution space makes exhaustive testing impractical. As the model's decision logic evolves during training, the possible range of each basis state is gradually compressed, akin to the bounded neuron outputs in DNNs. Therefore, we extract each probability range from the training data to profile the major and corner-case regions in QNNs.

Let $S=\{s_1, s_2, \ldots, s_{2^q}\}$ denote the set of basis states in a $q$-qubit circuit, $\mathcal{D}_{train}=\{x_1, x_2, \ldots, x_n\}$ denote the training data, and $\mathcal{D}_{test}=\{z_1, z_2, \ldots, z_m\}$ denote the test data, where $n$ and $m$ are the training and test size. A lower and upper bound for the probability range of each basis state, denoted as $l$ and $u$, depict the boundary between the major and corner-case regions. $P(s, z)$ represents the measured probability of the basis state $s$ for test input $z$.

\subsection{Profiling Process}
\label{sec:profiling}

Before coverage computation, we first need to profile major and corner-case regions for each basis state using a set of training data. Specifically, we feed data into the target QNN and collect measured probability distribution vectors, denoted as $R=\{r_1, r_2, \ldots, r_n\}$, where $r_j=(r_j^1, r_j^2, \ldots, r_j^{2^q})\in \mathbb{R}^{2^q}$ is obtained by $j$-th training data. For each basis state $s_i$ where $i=1, \ldots, 2^q$, its lower and upper bounds $l$ and $u$ are the minimum and maximum values occurring in the training data, respectively, defined as:
\begin{equation}
\label{eq:region}
\begin{split}
    l(s_i)=\min_{1 \le j \le n} r_j^i, \;
    u(s_i)=\max_{1 \le j \le n} r_j^i,\; \text{where}\; s_i\in S
\end{split}
\end{equation}
Consequently, the major and corner regions of the basis state $s_i$ are referred to as $[l(s_i), u(s_i)]$ and $[0, l(s_i))\cup (u(s_i), 1]$.

\subsection{Definitions}

To characterize how thoroughly a test suite explores the behavior space of a QNN from different perspectives, three coverage criteria are defined, namely $K$-cell State Coverage (KSC), State Corner Coverage (SCC) and Top-$k$ State Coverage (TSC). Their differences mainly lie in the granularity of covered units. KSC focuses on the coverage over major regions, SCC targets corner regions beyond those major regions, and TSC zooms in on a small set of impactful states.

\noindent
$\bullet$ \textbf{$K$-cell State Coverage.}
Since the probability range of each basis state is continuous, we discretize it into finite intervals for quantification. Specifically, for each basis state $s_i$, its major-region range $[l(s_i), u(s_i)]$ is divided into $k$ equal cells, denoted as $\mathcal{C}_s=\{c^1_s, c^2_s, \ldots, c^k_s\}$, where $c^i_s$ represents the $i$-th sub-range (cell), yielding $k*|S|$ cells in total. If $P(s, z) \in c^i_s$, we consider this cell covered by input $z$. $K$-cell State Coverage (KSC) is then defined as the ratio of cells covered by the test suite $\mathcal{D}_{test}$: 
\begin{equation} 
\label{eq:KSC} 
KSC(\mathcal{D}_{test}, k) = \frac{\sum_{s \in S}|\{c^i_s \mid \exists{z} \in \mathcal{D}_{test}, P(s, z) \in c^i_s\}|}{k*|S|} 
\end{equation} 

KSC reflects the extent to which the test suite covers major-region cells, providing a measure of how well primary QNN behaviors are tested.

\noindent
$\bullet$ \textbf{State Corner Coverage.}
The corner-case regions of a basis state are those outside the major one, i.e., $[0, l(s)) \cup (u(s), 1]$. Test inputs from distributions similar to the training data typically cover only a small fraction of corner-case regions. These regions are likely to reveal hidden erroneous behaviors~\cite{deepgauge}. To quantify this, we propose the State Corner Coverage (SCC), which measures the proportion of corner-case regions covered by the test suite and is computed as:
\begin{IEEEeqnarray}{rCl}
\label{eq:SCC}
SCC(\mathcal{D}_{test})
&=& \frac{1}{|S|}
\Bigl|\Bigl\{ s \in S \,\Bigm|\, \exists z \in \mathcal{D}_{test}, \nonumber\\
&&\qquad P(s,z)\in [0,l(s)) \cup (u(s),1] \Bigr\}\Bigr|
\end{IEEEeqnarray}

A higher SCC indicates that more previously unexplored regions have been covered, increasing the likelihood of uncovering potentially anomalous behaviors.

\noindent
$\bullet$ \textbf{Top-$k$ State Coverage.}
Transformed by quantum gates, the output states encode more abstract and input-specific features. For two inputs with similar distributions, their dominant states are likely to overlap. We refer to basis states with the largest $k$ probabilities as Top-$k$ states for a specific input. Beyond the collective statistics provided by KSC and SCC for behavioral space, Top-$k$ states aim to capture the comparative relationships between different states, highlighting a set of impactful states on model outputs. To this end, we define Top-$k$ states and Top-$k$ State Coverage (TSC) as: 
\begin{equation}
\begin{split}
    Top_k(z&) = \{t_1, \ldots, t_k\} \subseteq S, \\
\text{s.t. }  
\forall t_i \in Top_k(z), \; \forall & s \in S \setminus Top_k(z): \; 
P(t_i, z) \ge P(s, z)
\end{split}
\end{equation} 

\begin{equation}
    TSC(\mathcal{D}_{test}, k) = \frac{|\bigcup_{z\in \mathcal{D}_{test}}Top_k(z)|}{|S|}
\end{equation}

TSC measures the number of highly influential states that dominate the model's behaviors. A rapidly increasing TSC can indicate better test adequacy and suggest increased diversity in the model decision logic.
Thus, investigating the Top-$k$ states could potentially provide insights into the robustness of QNNs in future research.




\section{Research Questions}
\label{sec:rq}


\subsection{RQ1: \textbf{(Effectiveness)} Sensitivity to input diversity}

RQ1 focuses on evaluating the effectiveness of the coverage criteria as practical indicators of test adequacy for QNNs. We design the following two questions to examine whether and to what extent they can provide actionable signals for different test suites.

\begin{itemize}
    \item RQ1.1: Can coverage distinguish between test suites with different levels of diversity and overall quality?
    \item RQ1.2: What is the relationship between coverage criteria and the number of injected faults in test suites?
\end{itemize}

\subsection{RQ2: \textbf{(Scalability)} Application to QNNs of different circuit scales}

RQ2 focuses on assessing the scalability of coverage criteria as QNNs grow in circuit scale. The following question examines whether these criteria remain effective under increasing model sizes from two perspectives, namely the number of qubits and circuit depth, and whether they show potential applicability to larger quantum circuits in the future.

\begin{itemize}
    \item RQ2.1: How do the number of qubits and circuit depth affect the achievable coverage and the marginal gains from different test suites?
\end{itemize}

\subsection{RQ3: \textbf{(Robustness)} Impact of quantum-unique factors}

RQ3 focuses on investigating the robustness of coverage criteria under more realistic quantum-execution constraints. We design two questions to examine their sensitivity to measurement costs and their resilience to quantum noise.

\begin{itemize}
    \item RQ3.1: How does the number of measurement shots affect the reliability of coverage criteria?
    \item RQ3.2: Do coverage criteria remain effective under quantum noise?
\end{itemize}

\section{Experimental Setup}

We implement and evaluate coverage criteria \footnote{Replication package can be found in \url{https://anonymous.4open.science/r/CoverageCriteria-5047}.} using PennyLane 0.42~\cite{pennylane} and Pytorch 2.8~\cite{torch}. All experiments are conducted on systems equipped with Intel Xeon E5-1650 (6 cores, 32GB) and Ubuntu 22.04.

\subsection{Datasets and Models}
\label{sec:dataset}

\begin{table}[t]
\centering
\caption{Dataset and QNN architectures. \# gate refers to the total number of gates in a QNN.}
\label{tab:dataset}
\resizebox{\columnwidth}{!}{%
\begin{tabular}{llllll}
\hline
Dataset & Task & Target classes & QNN & \# gate & Acc (\%) \\ \hline
\multirow{5}{*}{MNIST} & \multirow{4}{*}{Binary classification} & \multirow{4}{*}{digits 0 and 1} & QCL & 150 & 99.52 \\
 &  &  & QCNN & 134 & 100 \\
 &  &  & HCQC & 64 & 100 \\
 &  &  & DRNN & 120 & 99.29 \\ \cline{2-6} 
 & Ternary classification & digits 0, 1 and 2 & QCL & 150 & 91.86 \\ \hline
\multirow{5}{*}{FashionMNIST} & \multirow{4}{*}{Binary classification} & \multirow{4}{*}{T-shirt and Trouser} & QCL & 150 & 92.5 \\
 &  &  & QCNN & 134 & 93 \\
 &  &  & HCQC & 64 & 94.25 \\
 &  &  & DRNN & 120 & 92.21 \\ \cline{2-6} 
 & Ternary classification & T-shirt, Trouser and Pullover & QCL & 150 & 89.67 \\ \hline
\end{tabular}
}
\end{table}

As for benchmarks, we choose two widely adopted datasets for image classification tasks. The MNIST dataset~\cite{mnist} contains 70,000 grayscale images of handwritten digits (0–9), each with a resolution of $28 \times 28$. The FashionMNIST dataset~\cite{fashion} has the same format but comprises images from ten categories of clothing items, such as T-shirts and trousers.

With respect to model architectures, besides two QNNs adopted in prior work \cite{quantest}, we also include two other QNNs with different block stacking and hierarchical structure strategies to formulate a more diverse evaluation.

\textbf{Quantum Circuit Learning (QCL)}~\cite{qcl} is a classical–quantum hybrid model designed to approximate nonlinear functions. QCL belongs to block-stacking structure and employs amplitude encoding. Its circuit depth\footnote{Here circuit depth refers to the repetition times of the building block.} is set to 5.

\textbf{Quantum Convolutional Neural Network (QCNN)}~\cite{qcnn} is inspired by classical CNNs, implementing convolution, pooling, and fully connected operations using quantum circuits. It is effective in mitigating the barren plateau problem \cite{quantumAI}. QCNN is hierarchical and employs amplitude encoding.

\textbf{Hierarchical Circuit Quantum Classifier (HCQC)}~\cite{hcqc} is another hierarchical model that reduces the circuit's degree of freedom as depth increases, using translationally stacked, invariant ansatz blocks. Here we use the ansatz U\_SO4\footnote{It consists of RY and CNOT gates with 6 parameters in total. The detailed structure can be found in \url{https://github.com/takh04/QCNN}.} with amplitude encoding. 

\textbf{Data Re-uploading Neural Network (DRNN)}~\cite{drnn} re-uploads data features as rotation angles to multiple qubits, combined with extra trainable parts. It addresses the limited expressivity of a single qubit with fewer quantum computational resources. DRNN adopts block-stacking structure and angle encoding, whose depth is 4.

Considering the time overhead, we adjust the image size for DRNN to $8\times8$ and configure it with 6 qubits, while the other QNNs are set to 8 qubits and images are downsampled to $16\times16$. We adopt 20\% of the original training and test datasets for all QNNs following \cite{hcqc}. QCLs trained for binary and ternary classification are referred to as QCL and QCL-3 respectively in following experiments.

To obtain reliable QNN outputs, we also need to estimate the minimum number of measurements $s$ to ensure an error bound $\epsilon$ with a confidence level of $\delta$. Let consider $P_i=\{X_i^{(1)},\dots,X_i^{(s)}\}$ as the outcomes of $s$ measurements on $i$-th qubit where $X_i^{(j)}$ is bounded and independent of each other. Ideal QNN outputs are defined as the expectation value of measurement results, denoted as $\mu$. The average estimate is $\hat{\mu}=\frac{1}{s}\sum_{j=1}^{s}X_i^{(j)}$. According to Hoeffding's inequality \cite{inequality}, for each qubit, $Pr(|\hat{\mu}-\mu|\geq \epsilon)\leq 2\text{exp}(-2s\epsilon^2)$. Then union bound is applied to all qubits and the total probability is bounded by $1-\delta$, producing $2q\text{exp}(-2s\epsilon^2)\leq 1-\delta$. Finally, we obtain $s\geq\frac{1}{2\epsilon^2}\text{log}(\frac{2q}{1-\delta})$.
Given that $q$ is 2 and 3 for binary and ternary classification respectively, we set $s=1000$ for a prediction to achieve a $\epsilon$ of 0.05 and a $\delta$ of 0.95.

\subsection{Test Suite Construction}
\label{sec:suite}

To examine whether coverage can quantify the adequacy of state exploration induced by different test suites, we construct suites with varying diversity and quality. As a reference, the original suite, denoted as \textit{Ori}, contains 100 inputs per class randomly sampled from the whole test datasets. We then categorize two groups of suites. 
\textbf{Strong} suites introduce more challenging or diverse inputs, including (1) \textit{LConf}, low-confidence inputs close to the decision boundary \cite{20emp,deepcrime}, (2) \textit{OOD}, out-of-distribution inputs drawn from unseen classes during training \cite{ood}, (3) \textit{Aug}, augmented inputs\footnote{Inputs are augmented via affine transformations, random cropping, brightness adjustment, blurring, sharpening and additive Gaussian and salt-and-pepper noise.}, and (4) \textit{Adv}, adversarial inputs produced by quantum FGSM algorithm\footnote{We estimate gradients using NES \cite{nes} for gradient-based attacks under finite shots. The step is configured as 50.} \cite{qaml}. For OOD and Adv, half of the original inputs are replaced with abnormal ones. These suites contain naturally or artificially hard-to-classify patterns and thus more likely to trigger model mistakes.

In contrast, \textbf{weak} suites exhibit reduced diversity or biased distributions, including (1) \textit{HConf}, high-confidence inputs far from the decision boundary, (2) \textit{Skewed}, class-imbalanced inputs with ratios of 10:1 for binary tasks and 10:1:1 for ternary tasks, and (3) \textit{Small}, a half-size suite compared to Ori. Except for Small, all suites have the same size as Ori.

\begin{table*}[t]
\centering
\caption{Coverage results (\%) on different test suites of MNIST. For the same QNN and criterion, \textbf{bold} indicates the highest value, and {\ul underling} indicates the lowest value.}
\label{tab:input_diversity_mnist}
\resizebox{0.85\textwidth}{!}{%
\begin{tabular}{@{}l|ccc|ccc|ccc|ccc|ccc@{}}
\toprule
\multicolumn{1}{c|}{} & \multicolumn{3}{c|}{QCL} & \multicolumn{3}{c|}{QCNN} & \multicolumn{3}{c|}{HCQC} & \multicolumn{3}{c|}{DRNN} & \multicolumn{3}{c}{QCL-3} \\ \cmidrule(l){2-16} 
\multicolumn{1}{c|}{\multirow{-2}{*}{Test suites}} & KSC & SCC & TSC & KSC & SCC & TSC & KSC & SCC & TSC & KSC & SCC & TSC & KSC & SCC & TSC \\ \midrule
Ori & 15.65 & 21.48 & 10.94 & 13.68 & 22.07 & 3.34 & 17.41 & 22.46 & 7.03 & 36.98 & 22.57 & 28.13 & 17.46 & 22.65 & 20.70 \\
HConf & \cellcolor[HTML]{EFEFEF}15.44 & \cellcolor[HTML]{EFEFEF}16.61 & \cellcolor[HTML]{EFEFEF}7.42 & \cellcolor[HTML]{EFEFEF}12.85 & \cellcolor[HTML]{EFEFEF}{\ul 12.31} & \cellcolor[HTML]{EFEFEF}{\ul 3.13} & \cellcolor[HTML]{EFEFEF}16.99 & \cellcolor[HTML]{EFEFEF}19.92 & \cellcolor[HTML]{EFEFEF}{\ul 6.25} & \cellcolor[HTML]{EFEFEF}34.39 & \cellcolor[HTML]{EFEFEF}16.41 & \cellcolor[HTML]{EFEFEF}25.00 & \cellcolor[HTML]{EFEFEF}17.27 & \cellcolor[HTML]{EFEFEF}22.85 & \cellcolor[HTML]{EFEFEF}14.06 \\
Skewed & \cellcolor[HTML]{EFEFEF}15.14 & \cellcolor[HTML]{EFEFEF}21.48 & \cellcolor[HTML]{EFEFEF}8.59 & \cellcolor[HTML]{EFEFEF}13.11 & \cellcolor[HTML]{EFEFEF}21.09 & \cellcolor[HTML]{EFEFEF}3.52 & \cellcolor[HTML]{EFEFEF}16.39 & \cellcolor[HTML]{EFEFEF}24.41 & \cellcolor[HTML]{EFEFEF}8.20 & \cellcolor[HTML]{EFEFEF}35.95 & \cellcolor[HTML]{EFEFEF}17.69 & \cellcolor[HTML]{EFEFEF}29.68 & \cellcolor[HTML]{EFEFEF}16.38 & \cellcolor[HTML]{EFEFEF}19.73 & \cellcolor[HTML]{EFEFEF}15.63 \\
Small & \cellcolor[HTML]{EFEFEF}{\ul 13.94} & \cellcolor[HTML]{EFEFEF}{\ul 14.26} & \cellcolor[HTML]{EFEFEF}{\ul 8.20} & \cellcolor[HTML]{EFEFEF}{\ul 11.52} & \cellcolor[HTML]{EFEFEF}12.70 & \cellcolor[HTML]{EFEFEF}3.52 & \cellcolor[HTML]{EFEFEF}{\ul 14.45} & \cellcolor[HTML]{EFEFEF}{\ul 14.06} & \cellcolor[HTML]{EFEFEF}{\ul 6.25} & \cellcolor[HTML]{EFEFEF}{\ul 28.46} & \cellcolor[HTML]{EFEFEF}{\ul 10.61} & \cellcolor[HTML]{EFEFEF}{\ul 18.75} & \cellcolor[HTML]{EFEFEF}{\ul 15.50} & \cellcolor[HTML]{EFEFEF}{\ul 15.62} & \cellcolor[HTML]{EFEFEF}{\ul 15.23} \\
LConf & 15.60 & 25.19 & 11.33 & \textbf{14.02} & 27.73 & \textbf{4.69} & \textbf{17.90} & 25.59 & 8.98 & 36.36 & 21.09 & 31.25 & 17.39 & 24.80 & 21.09 \\
OOD & 15.54 & 27.15 & 14.06 & 13.55 & 24.80 & \textbf{4.69} & 17.51 & 23.05 & 10.16 & 34.68 & 23.43 & 28.13 & 17.47 & 24.80 & 24.61 \\
Aug & \textbf{15.94} & \textbf{29.86} & 17.18 & 13.57 & 27.93 & \textbf{4.69} & 17.59 & \textbf{35.16} & 10.94 & 37.42 & 25.00 & 34.38 & 17.52 & 24.61 & 20.31 \\
Adv & 15.82 & 28.13 & \textbf{21.48} & 13.73 & \textbf{34.17} & 4.17 & 17.64 & 33.98 & \textbf{11.33} & \textbf{38.53} & \textbf{36.72} & \textbf{40.63} & \textbf{17.65} & \textbf{29.11} & \textbf{30.08} \\ \bottomrule
\end{tabular}%
}
\end{table*}

\begin{table*}[t]
\centering
\caption{Coverage results (\%) on different test suites of FashionMNIST.}
\label{tab:input_diversity_fashion}
\resizebox{0.85\textwidth}{!}{%
\begin{tabular}{@{}l|ccc|ccc|ccc|ccc|ccc@{}}
\toprule
\multicolumn{1}{c|}{} & \multicolumn{3}{c|}{QCL} & \multicolumn{3}{c|}{QCNN} & \multicolumn{3}{c|}{HCQC} & \multicolumn{3}{c|}{DRNN} & \multicolumn{3}{c}{QCL-3} \\ \cmidrule(l){2-16} 
\multicolumn{1}{c|}{\multirow{-2}{*}{Test suites}} & KSC & SCC & TSC & KSC & SCC & TSC & KSC & SCC & TSC & KSC & SCC & TSC & KSC & SCC & TSC \\ \midrule
Ori & 12.85 & 24.80 & 9.76 & 8.65 & 17.77 & 2.73 & 13.07 & 21.88 & 4.69 & 39.63 & 23.44 & 26.56 & 14.15 & 17.97 & 7.42 \\
HConf & \cellcolor[HTML]{EFEFEF}12.48 & \cellcolor[HTML]{EFEFEF}16.10 & \cellcolor[HTML]{EFEFEF}6.64 & \cellcolor[HTML]{EFEFEF}8.06 & \cellcolor[HTML]{EFEFEF}13.86 & \cellcolor[HTML]{EFEFEF}{\ul 2.34} & \cellcolor[HTML]{EFEFEF}12.83 & \cellcolor[HTML]{EFEFEF}20.90 & \cellcolor[HTML]{EFEFEF}4.30 & \cellcolor[HTML]{EFEFEF}36.42 & \cellcolor[HTML]{EFEFEF}16.40 & \cellcolor[HTML]{EFEFEF}18.75 & \cellcolor[HTML]{EFEFEF}13.78 & \cellcolor[HTML]{EFEFEF}16.80 & \cellcolor[HTML]{EFEFEF}5.86 \\
Skewed & \cellcolor[HTML]{EFEFEF}12.29 & \cellcolor[HTML]{EFEFEF}24.61 & \cellcolor[HTML]{EFEFEF}7.42 & \cellcolor[HTML]{EFEFEF}8.52 & \cellcolor[HTML]{EFEFEF}14.41 & \cellcolor[HTML]{EFEFEF}{\ul 2.34} & \cellcolor[HTML]{EFEFEF}12.80 & \cellcolor[HTML]{EFEFEF}22.07 & \cellcolor[HTML]{EFEFEF}5.47 & \cellcolor[HTML]{EFEFEF}35.28 & \cellcolor[HTML]{EFEFEF}17.96 & \cellcolor[HTML]{EFEFEF}25.00 & \cellcolor[HTML]{EFEFEF}{\ul 12.45} & \cellcolor[HTML]{EFEFEF}{\ul 11.52} & \cellcolor[HTML]{EFEFEF}{\ul 5.47} \\
Small & \cellcolor[HTML]{EFEFEF}{\ul 11.64} & \cellcolor[HTML]{EFEFEF}{\ul 16.02} & \cellcolor[HTML]{EFEFEF}{\ul 6.64} & \cellcolor[HTML]{EFEFEF}{\ul 7.08} & \cellcolor[HTML]{EFEFEF}{\ul 8.79} & \cellcolor[HTML]{EFEFEF}2.73 & \cellcolor[HTML]{EFEFEF}{\ul 11.39} & \cellcolor[HTML]{EFEFEF}{\ul 12.30} & \cellcolor[HTML]{EFEFEF}{\ul 3.52} & \cellcolor[HTML]{EFEFEF}{\ul 31.00} & \cellcolor[HTML]{EFEFEF}{\ul 10.94} & \cellcolor[HTML]{EFEFEF}{\ul 21.88} & \cellcolor[HTML]{EFEFEF}13.09 & \cellcolor[HTML]{EFEFEF}11.72 & \cellcolor[HTML]{EFEFEF}{\ul 5.47} \\
LConf & 12.79 & 25.00 & 8.20 & 8.72 & 22.27 & \textbf{2.73} & 13.24 & 21.68 & 5.47 & 40.14 & 21.88 & 26.56 & 14.08 & 17.18 & 6.64 \\
OOD & 13.20 & 34.77 & \textbf{11.33} & \textbf{9.39} & 31.25 & \textbf{4.68} & 13.67 & 35.94 & \textbf{8.20} & 38.91 & 34.38 & 46.88 & \textbf{14.76} & \textbf{33.98} & \textbf{9.38} \\
Aug & 13.16 & 31.83 & 8.20 & 9.08 & 27.73 & 2.73 & 13.50 & 35.94 & 7.03 & 40.44 & 30.47 & 35.93 & 14.33 & 24.02 & 7.03 \\
Adv & \textbf{13.24} & \textbf{33.79} & 9.38 & 9.20 & \textbf{39.69} & 3.13 & \textbf{13.86} & \textbf{38.89} & 6.64 & \textbf{40.98} & \textbf{42.98} & \textbf{62.50} & 14.61 & 32.42 & 8.20 \\ \bottomrule
\end{tabular}%
\vspace{-3mm}
}
\end{table*}

\subsection{Parameter Settings}

For profiling, we adopt training data following a standard distribution and uniformly sample 100 instances per class, with consideration of the dataset scale used for model training.
During evaluation, for KSC, $k$ denotes the number of cells within the major region of a basis state. For TSC, a lager Top-$k$ means more top-$k$ state candidates. We set $k=100$ and Top-$k=1$. The effect of parameter choice will be discussed in Section \ref{sec:ablation}. All the experiments are repeated for 10 times and average results are reported.

\section{Experimental Results}

\subsection{RQ1.1: Coverage results for different test suites}
\label{sec:rq1.1}

\black{In deep learning, the quality of a test suite is strongly influenced by its diversity, i.e., the extent to which the inputs span realistic variations and potential corner cases throughout the model lifecycle, including rare patterns, noise and distribution shifts. Such input diversity can expose failures that accuracy metric may overlook. Coverage criteria should serve as a proxy for test adequacy and quantify how thoroughly a test suite exercises model behaviors.}

\black{Here we compute coverage on the test suites constructed in Section \ref{sec:suite}. Taking Ori as the baseline, we expect weak suites to explore narrower regions and produce lower coverage. Strong suites are more likely to trigger distinct outputs and potential behaviors, yielding higher coverage.}

\black{According to results in Tables \ref{tab:input_diversity_mnist} and \ref{tab:input_diversity_fashion}, we have several findings.
\textbf{(1)} SCC and TSC are highly sensitive to input diversity. Specifically, compared to weak groups, SCCs on strong ones have increased by 59.47\% and 97.13\% (from 17.32\% to 27.62\% on MNIST, from 15.63\% to 30.80\% on FashionMNIST) respectively. TSCs have increased by 53.17\% and 70.10\%. The trend could be explained by the fact that LConf and Adv inputs lie closer to the decision boundary, either naturally or artificially constructed, are more likely to induce misclassifications and activate more corner-case regions as well as top states. OOD and Aug inputs introduce unseen features and yield outputs deviating from the major regions. Moreover, within the strong group, LConf inputs typically provide smaller gains than the other three suites, since they are directly drawn from the original test suite without additional perturbations.
\textbf{(2)} Compared to SCC and TSC, KSC varies less across suites. On strong groups, KSCs have increased by 9.47\% and 13.33\% on the two datasets respectively. A key reason is its finer granularity. Each major region is partitioned into $k* |S|$ cells whereas SCC considers only $2*|S|$ corner regions and TSC focuses on a few top states. Changes in major regions are weakened after being averaged over a large family of cells. Nevertheless, strong suites still generally obtain higher KSCs, indicating that they also promote broader state exploration within major regions.
\textbf{(3)} TSC exhibits subtler fluctuations, which depends on specific QNN design. The magnitude of TSC is more constrained, as top states can partly reflect the model's decision logic and tend to be relatively stable \cite{deepgauge}. In addition, the magnitude is architecture-dependent. For example, among 8-qubit QNNs and MNIST, QCNN and HCQC yield smaller TSCs than QCL. This is attributed to their hierarchical structures which progressively reduces the degree of freedom, narrowing the set of candidate states as top states. DRNN shows a larger magnitude due to fewer qubits.
\textbf{(4)} KSC and SCC are not mutually exclusive. We observe cases where both of them increase like MNIST and DRNN under Adv. This suggests that some inputs not only reach new corner regions but also are more dispersed across major regions. Conversely, both can decrease when inputs cover fewer corners and show a more concentrated coverage pattern within major regions. This phenomenon is due to the different granularities of KSC and SCC, and the fact that criteria focus on suite-level behaviors instead of individual inputs.
}

\black{Generally, these results highlight the necessity of including samples containing interesting patterns or lying close to decision boundary in improving state exploration.}

\begin{tcolorbox}[colback=black!1!white,colframe=black!40!white,
    left=0.66mm, right=0.66mm, top=0.66mm, bottom=0.66mm, boxsep=0.3mm, arc=2.5mm]
\textbf{Answer to RQ1.1:} \black{All three coverage criteria are sensitive to test suites of different levels of diversity, i.e., an increase of 9.47\%, 59.47\% and 53.17\% between strong and weak MNIST suites respectively.
Among them, SCC exhibits stronger sensitivity than KSC and TSC.}
\end{tcolorbox}

\subsection{RQ1.2: Relationship with the number of faults}
\label{sec:rq1.2}

\begin{figure*}[t]
    \centering
    \subfigure[QCL]{%
        \includegraphics[width=0.23\textwidth]{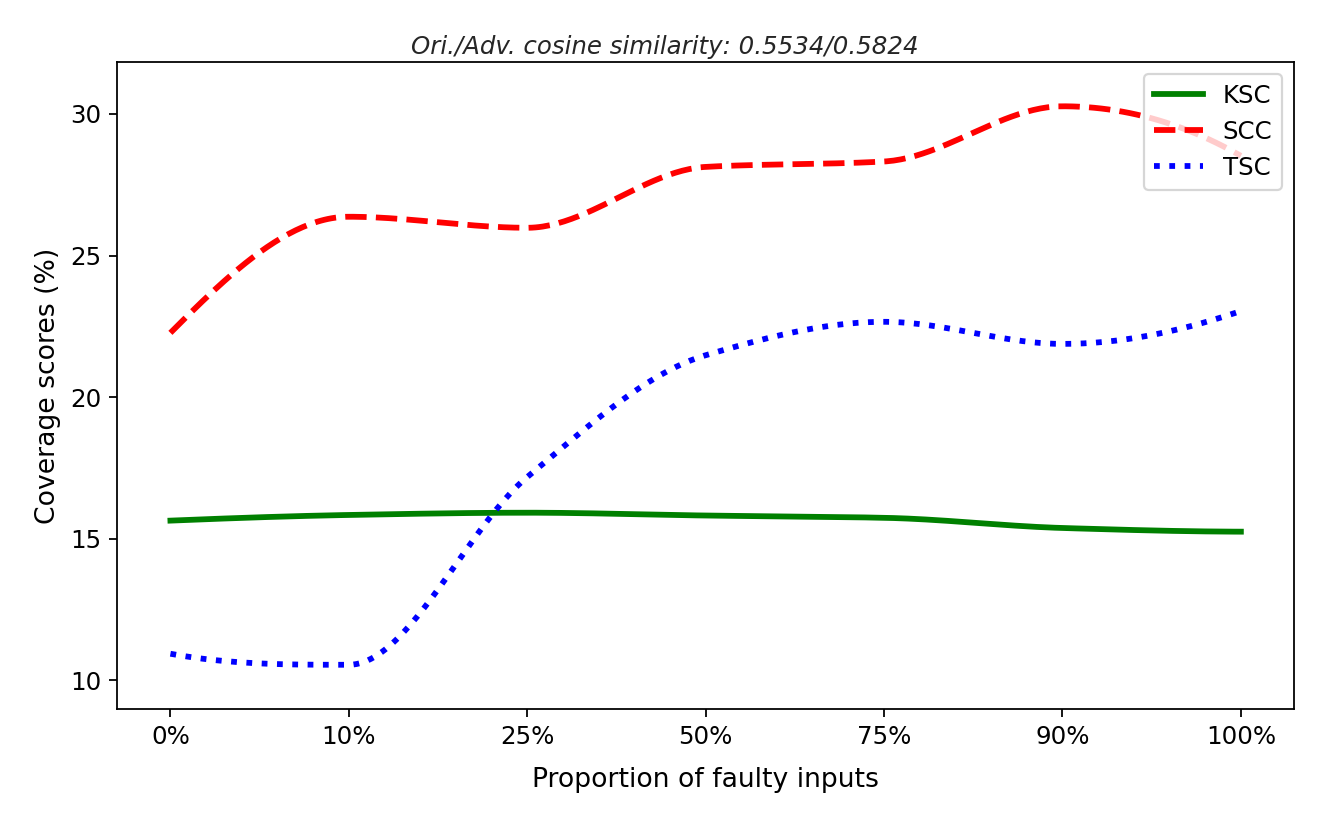}
        }
    \subfigure[QCNN]{%
        \includegraphics[width=0.23\textwidth]{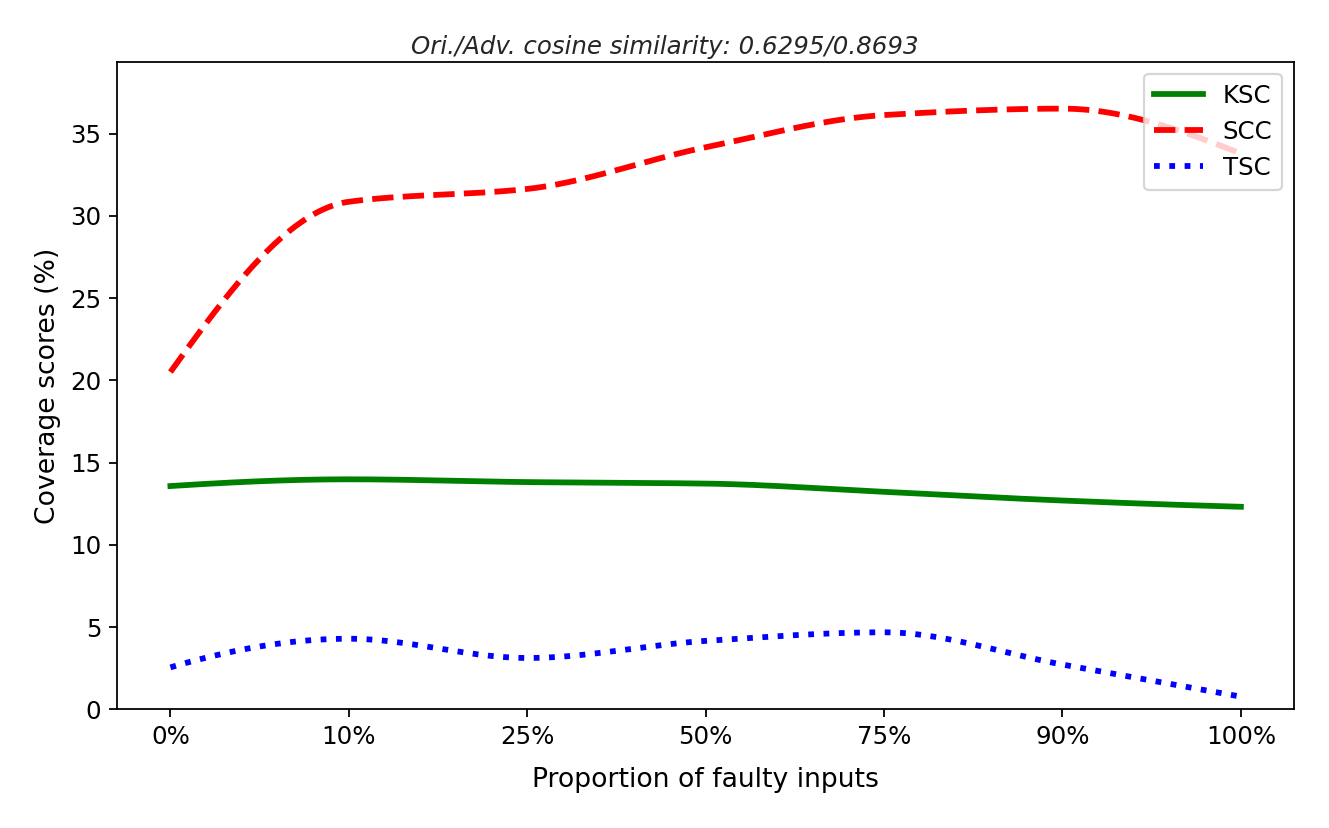}
        }
    \subfigure[HCQC]{%
        \includegraphics[width=0.23\textwidth]{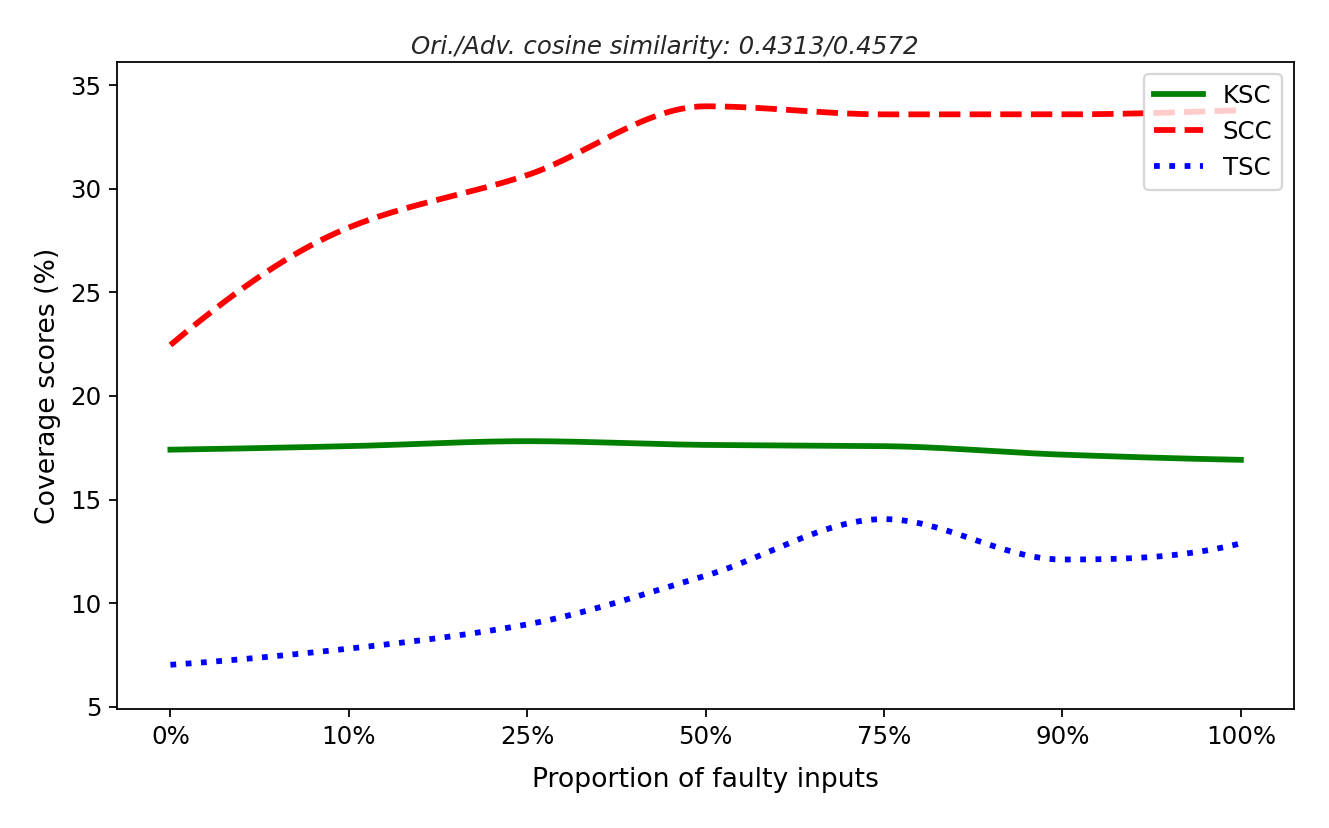}
        }
    \subfigure[DRNN]{%
        \includegraphics[width=0.23\textwidth]{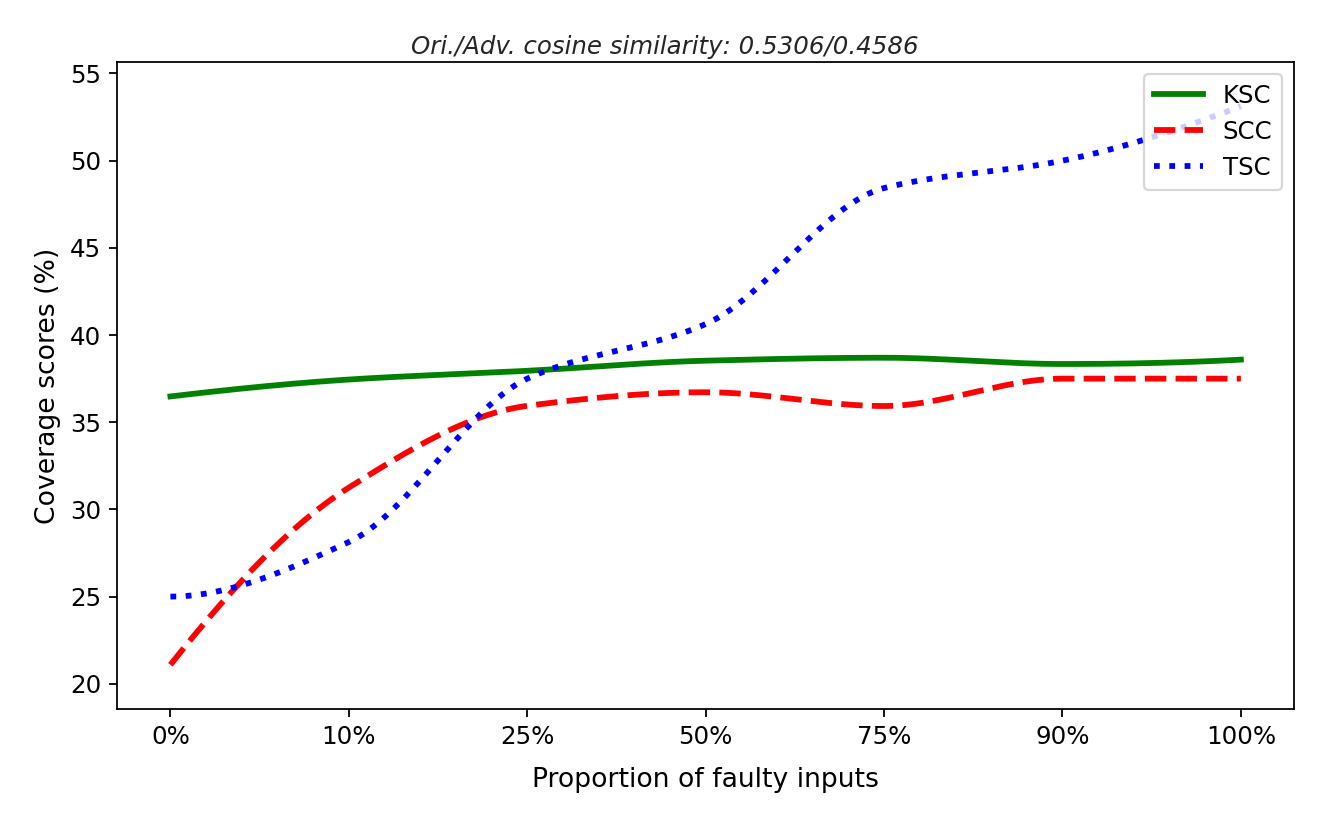}
        }
    \vspace{-2mm}
    \caption{Coverage results obtained by test suites containing different proportions of faults (MNIST).}
    \label{fig:proportion_fault_mnist}
\end{figure*}

\begin{figure*}[t]
    \centering
    \subfigure[QCL]{%
        \includegraphics[width=0.23\textwidth]{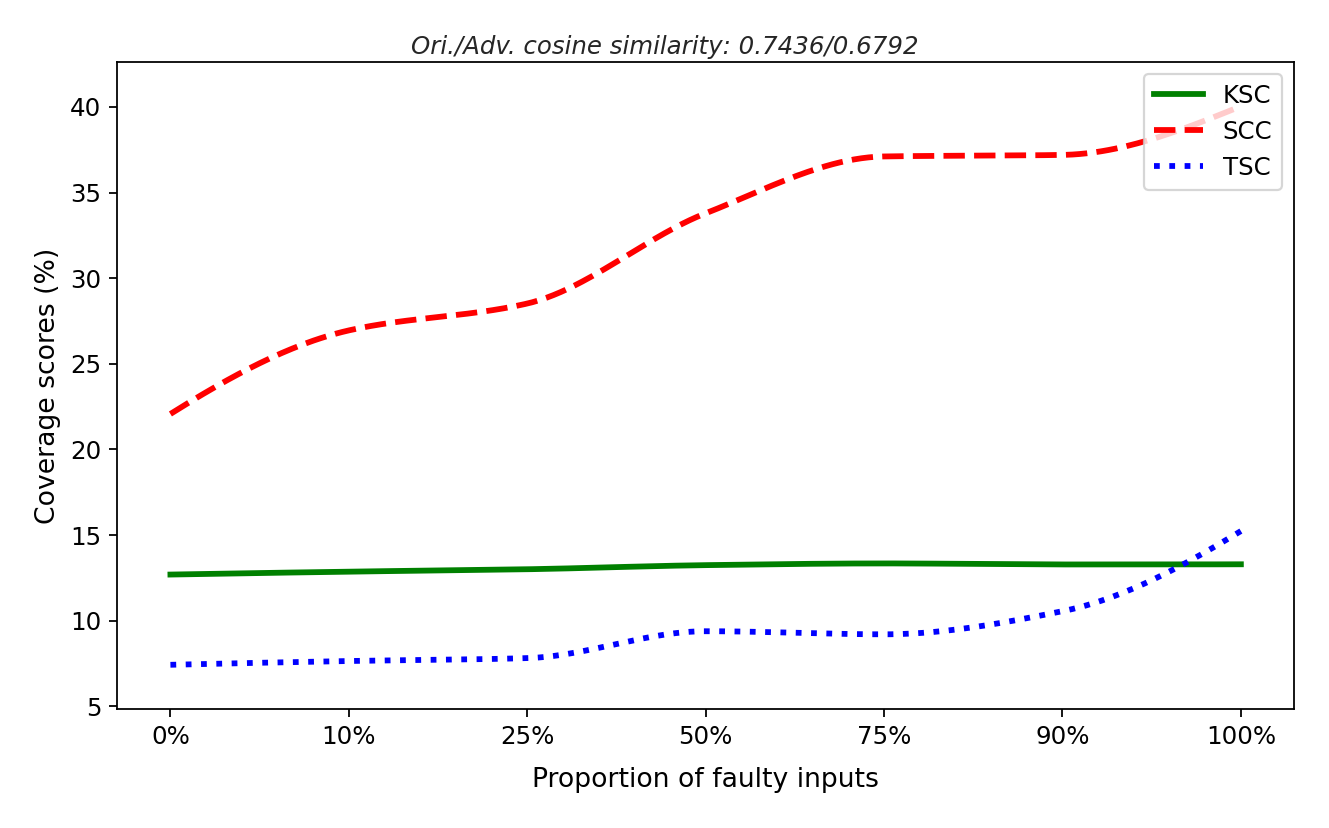}
        }
    \subfigure[QCNN]{%
        \includegraphics[width=0.23\textwidth]{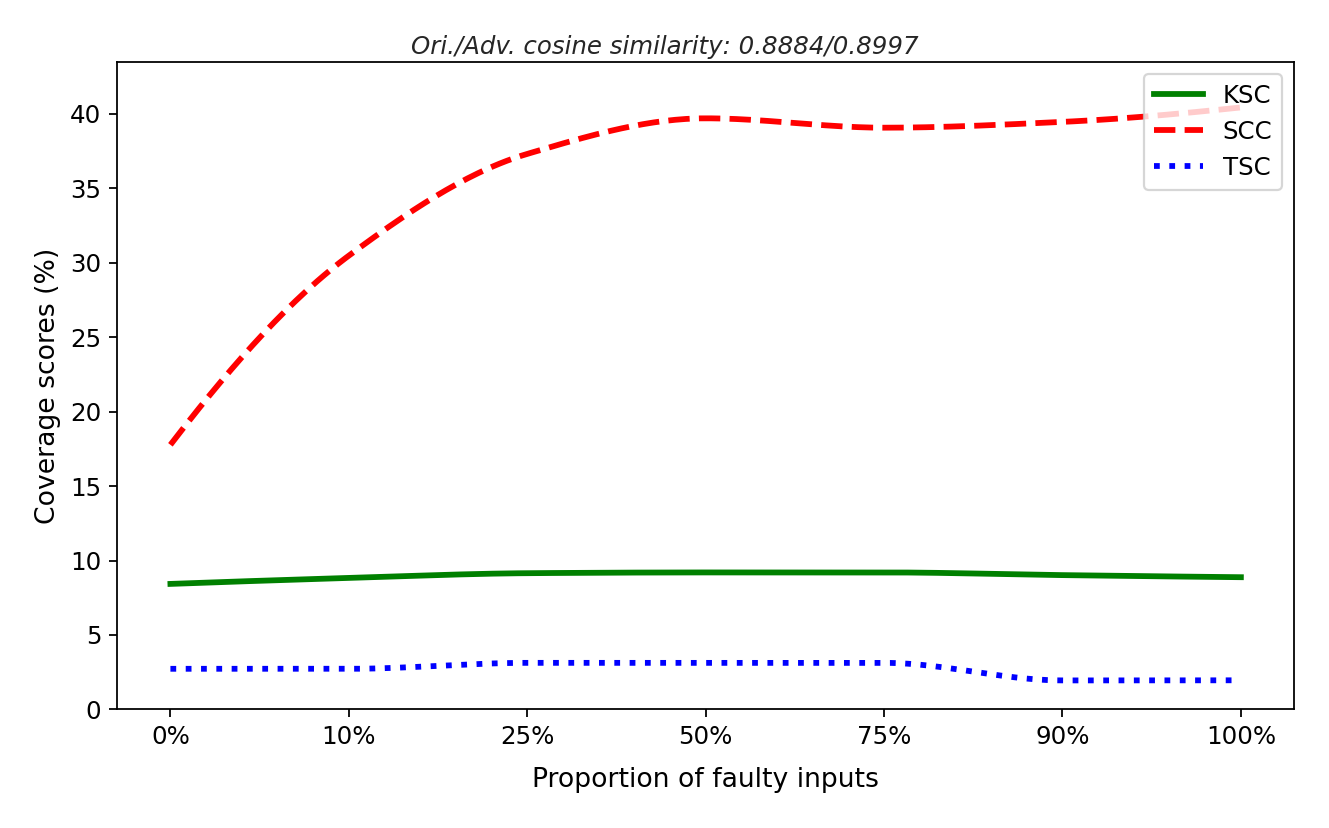}
        }
    \subfigure[HCQC]{%
        \includegraphics[width=0.23\textwidth]{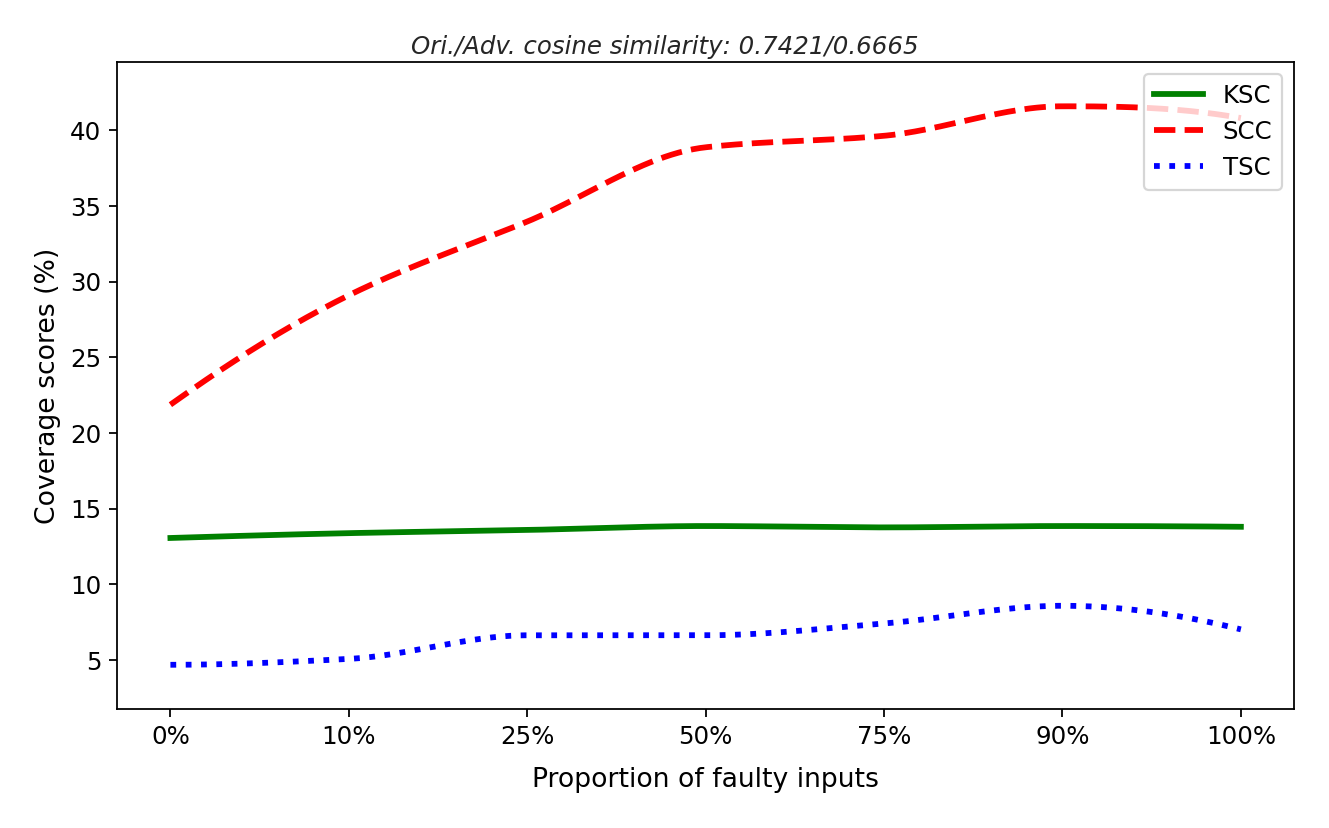}
        }
    \subfigure[DRNN]{%
        \includegraphics[width=0.23\textwidth]{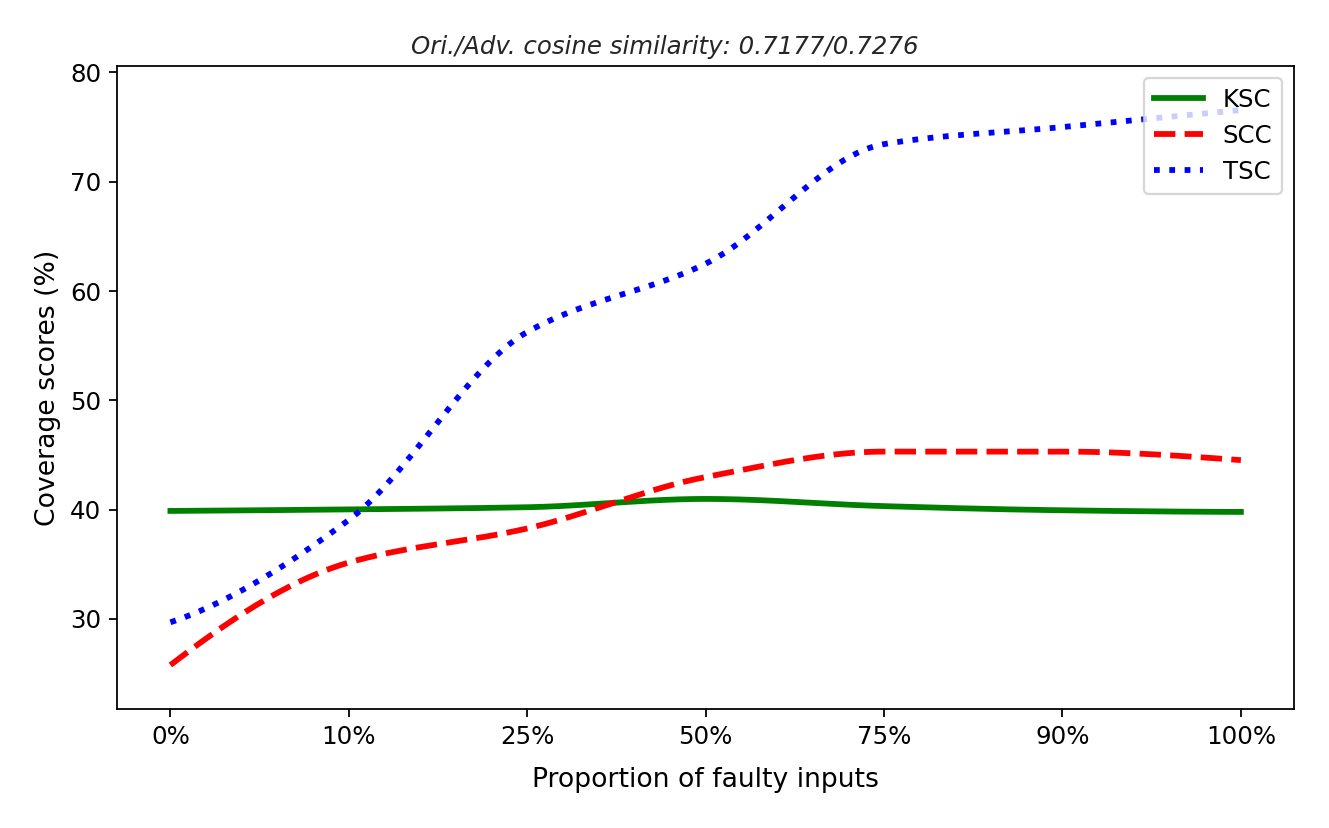}
        }
    \vspace{-2mm}
    \caption{Coverage results obtained by test suites containing different proportions of faults (FashionMNIST).}
    \label{fig:proportion_fault_fashion}
\end{figure*}

\black{
In classical software testing, even when new tests reveal program faults, they may exercise the same execution paths and trigger similar faults. In such cases, the structural coverage plateaus due to redundancy which stresses the diversity of faults instead of the number. In contrast to the code-level faults, faults in neural networks usually refer to failure-inducing inputs that occur erroneous behaviors. They are more diverse and their execution paths are inaccessible, making it difficult to determine the redundant faults. Therefore, in this RQ, we use the number of faults as a proxy for fault diversity and investigate whether including more faulty inputs can lead to higher coverage.
}

\black{
Considering that adversarial attacks are an efficient and mature technique for generating faulty inputs, we fix the test-suite size and vary the proportion of adversarial examples in the suite from 0\% to 100\%.
}

\black{
Results are depicted in \figurename \ref{fig:proportion_fault_mnist} and \ref{fig:proportion_fault_fashion}. We can observe that: 
\textbf{(1)} Coverage increases with faults up to the proportion of 75\%. Under these settings, the suites are mixtures of benign and adversarial samples, which jointly maintain coverage over major regions while expanding exploration of corner regions and inducing new top states.
\textbf{(2)} Coverage plateaus at larger proportions over 90\% where adversarial examples become dominant. This indicates substantial overlap among adversarial output patterns and limited improvement in coverage. On MNIST, except for DRNN, KSCs under 90\% and 100\% can be slightly lower than that of original suites. To figure out, we compute the cosine similarity of output probability vectors within original and adversarial suites respectively, as reported above each subfigure. A higher similarity within adversarial suites implies more redundant cells and hence lower KSCs. FGSM-like perturbations follow similar gradient directions for different inputs and cause concentration in output space. While DRNN exhibits the opposite trend, with lower adversarial similarity. This is attributed to its higher attack success rate and thus weaker robustness. On FashionMNIST, both original and adversarial suites show high similarity and are close to each other, resulting in smaller fluctuations of KSC.
\textbf{(3)} SCC remains positively correlated with the number of faults, consistent with the intuition that adversarial inputs tend to probe corner cases. The relationship between TSC and fault exposure is QNN-dependent. As mentioned before, for hierarchical structures like QCNN and HCQC, as the degree of freedom reduces, the top states become more concentrated, resulting in overlap among adversarial examples.
}

\begin{tcolorbox}[colback=black!1!white,colframe=black!40!white,
    left=0.66mm, right=0.66mm, top=0.66mm, bottom=0.66mm, boxsep=0.3mm, arc=2.5mm]
\textbf{Answer to RQ1.2:} \black{SCC shows a positive correlation with the number of faults, whereas KSC suffers more from the model robustness and redundancy among faults. TSC can be limited by hierarchical structures that reduce the diversity of impactful states.}
\end{tcolorbox}

\subsection{RQ2.1: Circuit scale}

\begin{table}[t]
\centering
\caption{Coverage results (\%) on MNIST and QCL of different numbers of qubits. ($\uparrow$) and ($\downarrow$) mean an increase and decrease relative to Ori respectively.}
\label{tab:qubit_number}
\resizebox{1.\columnwidth}{!}{%
\begin{tabular}{@{}c|c|l|ccc@{}}
\toprule
$\#$ qubits & Model Acc. & \multicolumn{1}{c|}{Test suites} & KSC  & SCC  & TSC \\ \midrule
\multirow{3}{*}{8} & \multirow{3}{*}{99.52\%} & Ori & 15.65 & 21.48 & 10.94 \\
 &  & Small & 15.44 (0.134$\downarrow$) & 16.61 (22.67$\downarrow$) & 7.42 (32.18$\downarrow$) \\
 &  & Aug & 15.60 (0.319$\downarrow$) & 25.19 (17.27$\uparrow$) & 11.33 (3.56$\uparrow$) \\ \midrule
\multirow{3}{*}{10} & \multirow{3}{*}{98.82\%} & Ori & 6.38 & 18.75 & 5.96 \\
 &  & Small & 5.86 (8.15$\downarrow$) & 12.94 (30.99$\downarrow$) & 4.01 (32.72$\downarrow$) \\
 &  & Aug & 6.41 (0.470$\uparrow$) & 21.63 (15.36$\uparrow$) & 6.93 (16.27$\uparrow$) \\ \midrule
\multirow{3}{*}{12} & \multirow{3}{*}{99.76\%} & Ori & 3.36 & 14.66 & 3.47 \\
 &  & Small & 3.13 (6.85$\downarrow$) & 8.61 (41.26$\downarrow$) & 1.86 (46.39$\downarrow$) \\
 &  & Aug & 3.37 (0.297$\uparrow$) & 15.86 (8.19$\uparrow$) & 3.49 (0.576$\uparrow$) \\ \bottomrule
\end{tabular}%
}
\end{table}

\begin{table}[t]
\centering
\caption{Coverage results (\%) on MNIST and QCL of different circuit depths.}
\label{tab:circuit_depth}
\resizebox{1.\columnwidth}{!}{%
\begin{tabular}{@{}c|c|l|ccc@{}}
\toprule
Depth & Model Acc. & \multicolumn{1}{c|}{Test suites} & KSC & SCC & TSC  \\ \midrule
\multirow{3}{*}{2} & \multirow{3}{*}{93.36\%} & Ori & 15.80 & 22.85 & 15.62 \\
 &  & HConf & 15.80 (0.0$-$) & 22.46 (1.71$\downarrow$) & 15.62 (0.0$-$) \\
 &  & LConf & 15.74 (0.379$\downarrow$) & 23.27 (1.84$\uparrow$) & 15.90 (1.79$\uparrow$) \\ \midrule
\multirow{3}{*}{5} & \multirow{3}{*}{99.52\%} & Ori & 15.65 & 21.48 & 10.94 \\
 &  & HConf & 15.44 (0.134$\downarrow$) & 16.61 (22.67$\downarrow$) & 7.42 (32.18$\downarrow$) \\
 &  & LConf & 15.60 (0.319$\downarrow$) & 25.19 (17.27$\uparrow$) & 11.33 (3.56$\uparrow$) \\ \midrule
\multirow{3}{*}{10} & \multirow{3}{*}{99.29\%} & Ori & 16.16 & 23.24 & 15.63 \\
 &  & HConf & 15.49 (4.15$\downarrow$) & 14.06 (39.50$\downarrow$) & 7.82 (49.97$\downarrow$) \\
 &  & LConf & 16.29 (0.804$\uparrow$) & 25.20 (8.43$\uparrow$) & 17.19 (9.98$\uparrow$) \\ \bottomrule
\end{tabular}%
}
\end{table}

\black{
Scalability in quantum software testing concerns whether a testing technique remains effective as quantum circuits grow in complexity, which is typically characterized by the number of qubits, circuit depth and gate count \cite{mutationoperator}. Given the future growth of circuit complexity and available qubits, this RQ investigates the scalability of coverage criteria to different scales of QNNs. Specifically, on MNIST, we vary the number of qubits from 8 (default setting), 10 to 12, and circuit depth from 2, 5 (default setting) to 10 for QCL.
}

\black{
Observations include:
\textbf{(1)} From Table \ref{tab:qubit_number}, the overall magnitude of coverage decreases as the number of qubits increase. This is intuitive since the number of basis states grows exponentially. Also, it becomes harder for strong suites to improve coverage. The gains under Aug tend to shrink for more-qubit QNNs, while the degradation under Small becomes more pronounced, which indicates that it becomes more challenging to produce distinct output patterns in a grown state space under a generally stable model performance. 
\textbf{(2)} From Table \ref{tab:circuit_depth}, the sensitivity of coverage to test diversity generally increases with circuit depth. Deeper circuits usually contain more complicated gate operations and entangling structures, contributing to a model with stronger expressivity \cite{qleet}. Such models can generate more possible output states, as reflected in more obvious relative changes of coverage. In contrast, shallow circuits have limited expressive capacity, causing coverage closer between different suites. However, prior work \cite{sim2019expressibility} has also pointed out that expressivity may saturate beyond certain number of depths or layers.
}

\begin{tcolorbox}[colback=black!1!white,colframe=black!40!white,
    left=0.66mm, right=0.66mm, top=0.66mm, bottom=0.66mm, boxsep=0.3mm, arc=2.5mm]
\textbf{Answer to RQ2.1:} \black{The magnitude of coverage results decreases on QNNs with more qubits, in correspondence with more basis states. While on deeper circuits with stronger expressivity, criteria exhibit larger relative changes.}
\end{tcolorbox}

\subsection{RQ3.1: Number of shots}

\begin{figure}[t]
  \centering
  \subfigure[Comparison with ideal results]{%
        \includegraphics[width=0.23\textwidth]{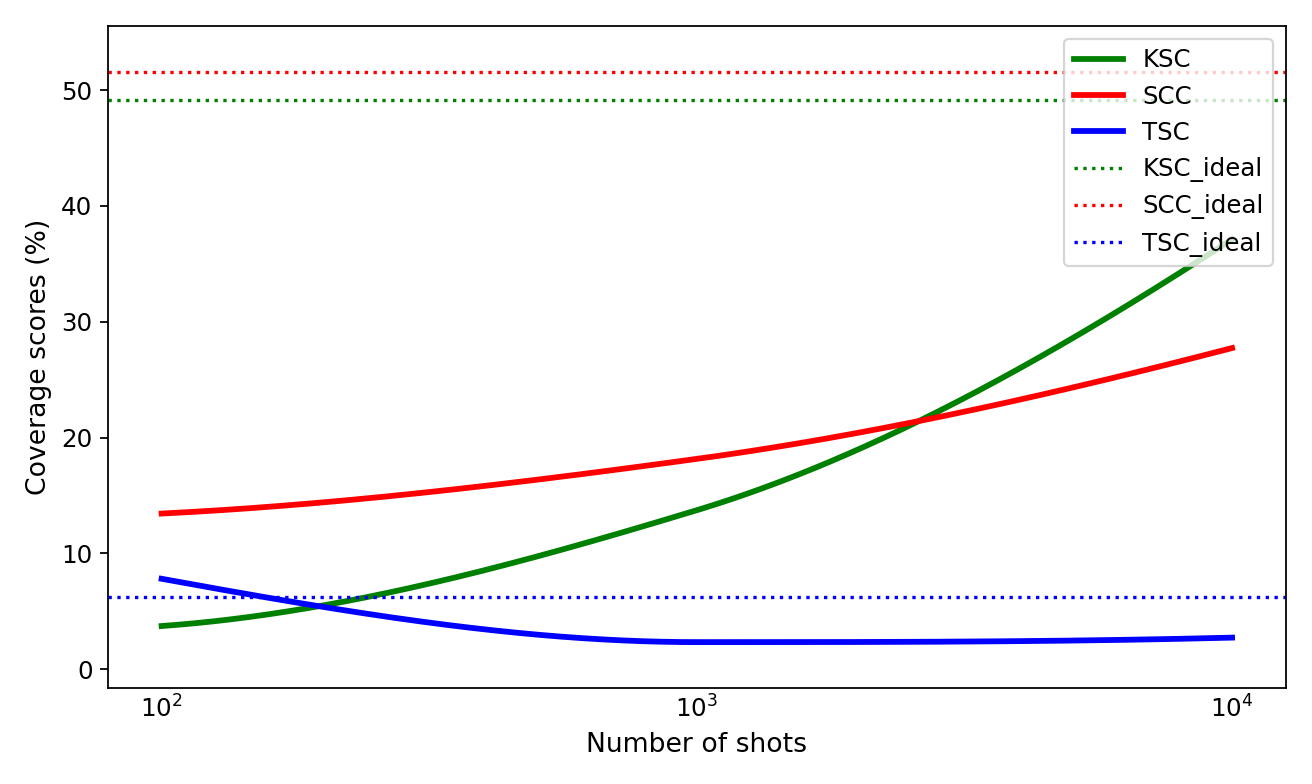}
        \label{fig:shot_ideal}
        }
  \subfigure[Results on Ori and OOD suites]{%
        \includegraphics[width=0.23\textwidth]{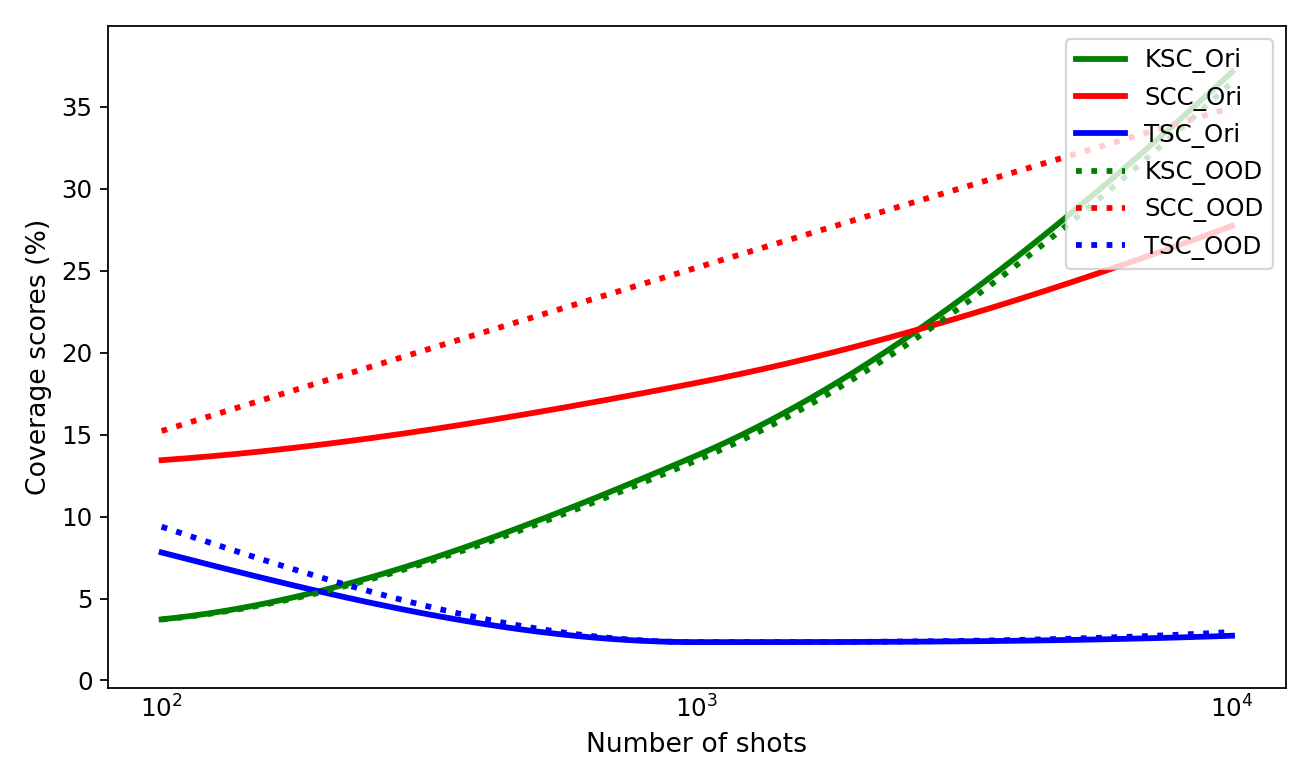}
        \label{fig:shot_diversity}
        }
    \caption{The effect of the number of shots on coverage results computed on MNIST and QCNN.}
    \label{fig:shot}
\end{figure}

\black{
Insufficient shots can introduce strong randomness to QNN outputs, thus threatening the testing accuracy. Since a large shot budget may be costly in practice, this RQ aims to study the performance of coverage criteria with different shots. Specifically, by setting the parameter \texttt{shots} of function \texttt{qml.device} in PennyLane as $10^2$, $10^3$ (default setting for other experiments) and $10^4$, QCNN is retrained on MNIST respectively. For comparison, we also include the model trained under infinite shots, i.e., an ideal baseline.
}

\black{
From \figurename \ref{fig:shot_ideal}, we can find that:
\textbf{(1)} Coverage results deviate from the ideal ones at low shots like $10^2$, indicating the impact of measurement randomness. For KSC and SCC, regions with low probabilities are easy to be missed. In contrast, TSC exhibits a decreasing trend. Under $10^2$ shots, it appears higher because some originally non-top states are mistaken as top ones due to insufficient measurement. As shots increase, coverage results approach their ideal cases gradually. 
\textbf{(2)} Coverage criteria remain sensitive to input diversity as shown in \figurename \ref{fig:shot_diversity}. The sensitivity of KSC and SCC becomes more stable with larger shot budgets, indicating the necessity to ensure stable model outputs and reliable testing results in practical scenarios.
}

\begin{tcolorbox}[colback=black!1!white,colframe=black!40!white,
    left=0.66mm, right=0.66mm, top=0.66mm, bottom=0.66mm, boxsep=0.3mm, arc=2.5mm]
\textbf{Answer to RQ3.1:} \black{A larger shot budget can improve the precision of coverage estimation. Even with a few shots, coverage criteria can still be sensitive to input diversity.}
\end{tcolorbox}

\subsection{RQ3.2: Under quantum noise}

\begin{table}[t]
\centering
\caption{Coverage results (\%) on noiseless and noisy HCQCs with U\_TTN ansatz.}
\label{tab:noise}
\resizebox{0.75\columnwidth}{!}{%
\begin{tabular}{@{}l|ccc|ccc@{}}
\toprule
 & \multicolumn{3}{c|}{Noiseless model} & \multicolumn{3}{c}{Noisy model} \\ \cmidrule(l){2-7} 
\multirow{-2}{*}{Test suites} & KSC & SCC & TSC & KSC & SCC & TSC \\ \midrule
Ori & 12.57 & 15.23 & 3.13 & 12.63 & 15.04 & 2.34 \\
HConf & \cellcolor[HTML]{EFEFEF}11.83 & \cellcolor[HTML]{EFEFEF}10.74 & \cellcolor[HTML]{EFEFEF}{\ul 1.95} & \cellcolor[HTML]{EFEFEF}12.20 & \cellcolor[HTML]{EFEFEF}11.33 & \cellcolor[HTML]{EFEFEF}2.34 \\
Skewed & \cellcolor[HTML]{EFEFEF}12.55 & \cellcolor[HTML]{EFEFEF}15.43 & \cellcolor[HTML]{EFEFEF}2.34 & \cellcolor[HTML]{EFEFEF}12.28 & \cellcolor[HTML]{EFEFEF}15.04 & \cellcolor[HTML]{EFEFEF}2.34 \\
Small & \cellcolor[HTML]{EFEFEF}{\ul 9.72} & \cellcolor[HTML]{EFEFEF}{\ul 7.42} & \cellcolor[HTML]{EFEFEF}{\ul 1.95} & \cellcolor[HTML]{EFEFEF}{\ul 9.49} & \cellcolor[HTML]{EFEFEF}{\ul 9.96} & \cellcolor[HTML]{EFEFEF}{\ul 1.95} \\
LConf & 12.89 & 15.43 & 3.13 & \textbf{12.76} & 19.14 & \textbf{2.73} \\
OOD & 12.58 & 20.12 & 3.52 & 12.66 & 20.31 & 2.34 \\
Aug & 12.39 & 19.14 & \textbf{3.91} & 12.36 & 20.51 & \textbf{2.73} \\
Adv & \textbf{12.94} & \textbf{24.41} & 3.13 & 12.59 & \textbf{25.59} & \textbf{2.73} \\ \bottomrule
\end{tabular}%
}
\end{table}

\black{
Real quantum systems suffer from unwanted interactions with the outside world, showing up as noise including gate-level errors and decoherence noise during circuit execution \cite{nielsen2010quantum}. These noise can inevitably affect the transformation of quantum states and final measurement results, impairing the accuracy of testing. In this RQ, we explore the robustness of coverage criteria to quantum noise.
}

\black{
Due to limited access to real quantum hardware, we simulate noisy execution using \texttt{mixed-state} device in PennyLane and a set of representative noise channels, including  depolarizing, amplitude damping, phase damping, bit flip, phase flip, crosstalk and thermal relaxation. We model these channels whose parameter choices follow Qiskit documents\footnote{Please refer to \url{https://quantum.cloud.ibm.com/docs/en/guides/get-qpu-information} and \url{https://qiskit.github.io/qiskit-aer/tutorials/3_building_noise_models.html}.}. To reduce time overhead, we train a smaller-scale HCQC with ansatz U\_TTN \footnote{Its scale is smaller than U\_SO4, composed of RY and CNOT gates with 2 parameters in total.} under both noisy and noiseless settings.
}

\black{As in Table \ref{tab:noise}, noise may shift the absolute coverage results, but the relative trends are largely preserved. A key reason is that the noise acts as a moderate perturbation at the output-distribution level. KSC and SCC depend on region-level behaviors at a finer granularity, making most executions locate in the same region. Additionally, coverage is computed from estimated probabilities over repeated measurements, which helps mitigate the stochastic fluctuations of noise.}


\black{TSC is most sensitive, showing compressed values and reduced discriminative power. In particular, under LConf, Aug and Adv, TSC yields the same score, whereas their top states are expected to differ on the noiseless model. This is due to that it depends on a small set of highly influential states, which can be easily influenced by noisy perturbations.}

\begin{tcolorbox}[colback=black!1!white,colframe=black!40!white,
    left=0.66mm, right=0.66mm, top=0.66mm, bottom=0.66mm, boxsep=0.3mm, arc=2.5mm]
\textbf{Answer to RQ3.2:} \black{Despite the stochastic influence of noise on circuit execution, the outputs can remain discriminative and be reflected in corresponding coverage criteria.}
\end{tcolorbox}

\subsection{Ablation Study}
\label{sec:ablation}

\paragraph{Effect of parameter choice}

\begin{figure}[t]
  \centering
  \subfigure[Different $k$ for KSC]{%
        \includegraphics[width=0.23\textwidth]{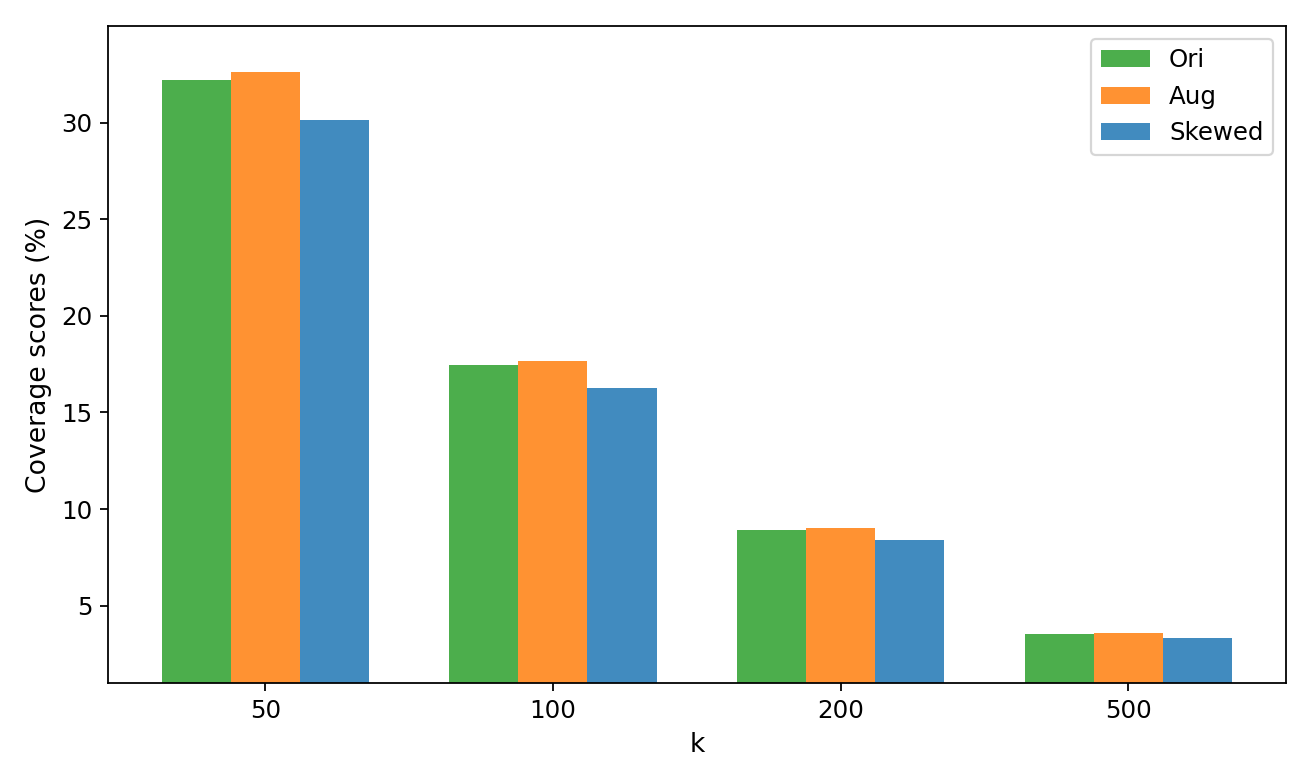}
        }
  \subfigure[Different Top-$k$ for TSC]{%
        \includegraphics[width=0.23\textwidth]{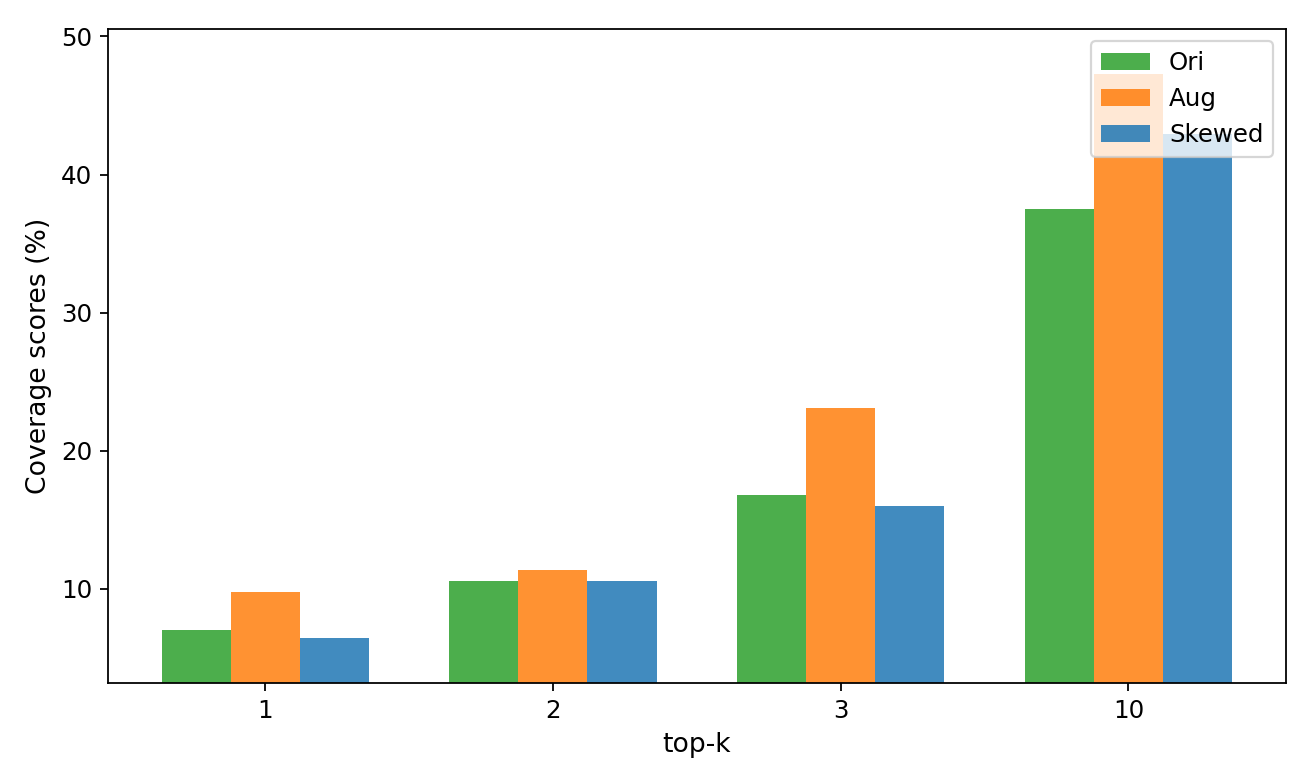}
        }
    \caption{Coverage results under different parameter settings computed on MNIST and HCQC.}
    \label{fig:param_choice}
\end{figure}

\black{
To study the impact of hyperparameters including $k$ in KSC and Top-$k$ in TSC, we conduct an ablation analysis with different settings and recomputing coverage for Ori, Skewed and Aug suites.
Results in \figurename \ref{fig:param_choice} are intuitive. Increasing $k$ partitions the major region into more cells, leading to a finer granularity and a lower KSC. A lager Top-$k$ can allow more states to rank as candidates and cause a larger TSC. 
Despite these numerical shifts, KSC and TSC remain responsive to differences between test suites. 
Moreover, for KSC, an excessive $k$ may introduce too many cells and impact its sensitivity. For example, the relative increase between Ori and Aug when $k=50$ is 1.272\% compared to 1.012\% when $k=500$.
}

\paragraph{Effect of training data distribution}

\begin{table}[t]
\centering
\caption{Effects of training data distribution on KSC and SCC (\%). Results are computed on the original MNIST suite. $\downarrow$ and $\uparrow$ mean a decrease and increase compared to results under Ori distribution.}
\label{tab:train_dis}
\resizebox{0.95\columnwidth}{!}{%
\begin{tabular}{@{}l|l|cccccccc@{}}
\toprule
\multirow{2}{*}{QNN} & \multirow{2}{*}{Criteria} & \multicolumn{8}{c}{Training data distribution} \\ \cmidrule(l){3-10} 
 &  & Ori & HConf & Skewed & Small & LConf & OOD & Aug & Adv \\ \midrule
\multirow{2}{*}{QCL} & KSC & 15.65 & 14.94 $\downarrow$ & 15.25 $\downarrow$ & 14.96 $\downarrow$ & 15.28 $\downarrow$ & 15.45 $\downarrow$ & 15.75 $\uparrow$ & 15.89 $\uparrow$ \\
 & SCC & 21.48 & 31.64 $\uparrow$ & 25.39 $\uparrow$ & 32.23 $\uparrow$ & 20.12 $\downarrow$ & 17.19 $\downarrow$ & 18.16 $\downarrow$ & 18.75 $\downarrow$ \\ \midrule
\multirow{2}{*}{QCNN} & KSC & 13.68 & 12.38 $\downarrow$ & 12.63 $\downarrow$ & 13.17 $\downarrow$ & 13.19 $\downarrow$ & 13.18 $\downarrow$ & 13.51 $\downarrow$ & 13.73 $\uparrow$ \\
 & SCC & 22.07 & 40.82 $\uparrow$ & 23.24 $\uparrow$ & 29.29 $\uparrow$ & 20.31 $\downarrow$ & 18.16 $\downarrow$ & 13.87 $\downarrow$ & 15.82 $\downarrow$ \\ \midrule
\multirow{2}{*}{HCQC} & KSC & 17.57 & 16.71 $\downarrow$ & 15.18 $\downarrow$ & 17.11 $\downarrow$ & 16.04 $\downarrow$ & 16.83 $\downarrow$ & 17.63 $\uparrow$ & 18.06 $\uparrow$ \\
 & SCC & 25.00 & 31.64 $\uparrow$ & 27.15 $\uparrow$ & 30.27 $\uparrow$ & 23.24 $\downarrow$ & 23.64 $\downarrow$ & 17.57 $\downarrow$ & 16.80 $\downarrow$ \\ \midrule
\multirow{2}{*}{DRNN} & KSC & 36.98 & 31.55 $\downarrow$ & 34.72 $\downarrow$ & 37.17 $\uparrow$ & 35.79 $\downarrow$ & 35.73 $\downarrow$ & 36.84 $\downarrow$ & 34.20 $\downarrow$ \\
 & SCC & 22.57 & 34.37 $\uparrow$ & 29.68 $\uparrow$ & 28.91 $\uparrow$ & 25.00 $\uparrow$ & 21.88 $\downarrow$ & 17.97 $\downarrow$ & 6.25 $\downarrow$ \\ \midrule
\multirow{2}{*}{QCL-3} & KSC & 17.46 & 16.86 $\downarrow$ & 15.61 $\downarrow$ & 16.89 $\downarrow$ & 16.78 $\downarrow$ & 17.29 $\downarrow$ & 17.48 $\uparrow$ & 17.34 $\downarrow$ \\
 & SCC & 22.65 & 29.10 $\uparrow$ & 31.25 $\uparrow$ & 33.01 $\uparrow$ & 20.81 $\downarrow$ & 19.14 $\downarrow$ & 20.31 $\downarrow$ & 18.75 $\downarrow$ \\ \bottomrule
\end{tabular}%
}
\end{table}

\black{
The profiling process is fundamental during which training data determines the division between major and corner-case regions. However, if the profiling data are of low quality or follow an unrepresentative distribution, e.g., being biased towards some classes, the resulting partition can be distorted and undermine the reliability of coverage criteria. To simulate such cases, we construct alternative training data following Section \ref{sec:suite} and repeat the profiling process. Intuitively, weak profiling data should yield smaller major regions and larger corner ones, whereas the opposite holds for strong data.
}

\black{
Coverage results on Ori suites are reported in Table \ref{tab:train_dis}.
SCC has shown different patterns under different training data distributions. Under weak data settings, SCC witnesses noticeable increase, consistent with expanded corner regions caused by smaller major regions. In contrast, SCC decreases under strong settings due to narrower corner regions.
We also find that KSC is less affected. Although the overall extent of major regions has changed, KSC further partitions them into $k$ cells per state. The averaged cell-level changes tend to be subtle and result in unobvious fluctuations in the final KSC.
}

\section{Discussion}

\subsection{Granularities of Coverage Criteria}

\black{
From RQ1, the criteria exhibit different sensitivities to input diversity. SCC and TSC are generally more discriminative than KSC. A main reason is that KSC uses a hard rule for covering a major cell where a cell is covered once any input has probability located in its range, regardless of coverage difficulty or frequency. Future work can consider softer rules, such as incorporating coverage frequency or weighted computation.
For SCC, non-major regions are aggregated into two coarse blocks (upper and lower). A possible extension is to further divide these regions to increase granularity. 
Moreover, criteria are defined at test-suite level, targeting cumulative adequacy over a suite instead of individual inputs. To fill this gap, input-level notions such as surprise adequacy \cite{surprise}, could be incorporated to quantify per-input contributions.
}

\subsection{Application in Guiding Test Generation}

\black{
The results of RQ1.2 suggest that coverage criteria can correlate with the number of faults, and SCC further exhibits a positive association. However, as observed in the coverage-guided DNN testing \cite{deephunter}, whether increasing coverage can help generate more fault-inducing inputs remains an open question \cite{zhang2019investigation}. This is because higher coverage only indicates newly explored regions while does not necessarily imply the presence of faults or proximity to the decision boundary. For example, data augmentation can diversify features without the goal of misleading models. 
In QP testing, results in \cite{quratest} do not demonstrate a positive correlation between coverage and mutation scores as well.
Nevertheless, given the complicated and unpredictable behaviors of neural networks, future work can explore how coverage-guided test generation influences properties beyond failure discovery, such as generalization, correctness, and unfairness.
}

\subsection{Challenges for Testing Large-scale QNNs}

\black{As discussed in RQ2.1, large-scale circuits face high-dimensional output spaces, which make thorough testing more difficult. Meanwhile, obtaining QNN outputs incurs higher resource and runtime costs. In such cases, enlarging the suite size might be an efficient strategy to improve coverage but bring about higher costs in practice. Another option is to include a moderate portion of fault-inducing inputs like adversarial examples. From the perspective of model expressivity, coverage criteria may also serve as an auxiliary indicator of expressivity saturation, helping identify the boundaries and enhance circuit structures.}

\subsection{Testing Under Realistic Quantum Execution}

\black{
Due to the stochasticity of quantum measurement, the shot budget must be considered to ensure reliable testing. On the one hand, there is still no clear theoretical guidance on how many shots are enough to yield stable outputs. On the other hand, practical costs including access to quantum resources have limited the feasibility of using a tremendous number of shots. To compensate for this, at current phase of testing QNNs, more experimental repetitions could be included to improve the statistical significance of evaluation.
}

\black{
Considering the inherent noise in quantum hardware, testing techniques should remain effective on real devices. Nevertheless, simulator-based evaluation cannot be denied arbitrarily as it has been widely adopted in QML research. When the goal is to reveal the model-intrinsic faults rather than to study the environmental effects of noise, simulators provide a reasonable and reproducible setting for early-stage practices. Especially, access to real devices is expensive and outcomes can be heavily affected by hardware-specific factors like noise rates and qubit connectivity, reducing testing efficiency and accuracy. In the long term, progress towards fault-tolerant quantum computers~\cite{preskill2018quantum} indicates that post-NISQ devices may demonstrate practical applicability and show relaxed constraints, reducing the mismatch between idealized assumptions and practical deployment.
Overall, since noise is still unavoidable in the near term, future work should explicitly assess the impact of quantum noise even using simulated noise.
}

\section{Threats to validity}

\textit{External validity.}
\black{
Due to the current scale and accuracy of QNNs, our study focuses on relatively small benchmarks such as MNIST and implements only binary and ternary classification tasks. 
Although multiple representative QNN architectures are evaluated, future work may consider additional encoding schemes and circuit designs that could affect coverage results.
Moreover, our evaluation focuses on image classification tasks following existing QNN studies, while the proposed coverage criteria could be extended to other tasks such as sequential data processing and generative models.
In addition, noise is simulated and injected with a fixed probability, which differs from real hardware noise that is device-specific and unpredictable.
Our study on circuit complexity is also limited to several settings of qubit numbers and circuit depths. The experimental scale could be further enhanced by applying the criteria to larger-scale QNNs as simulators and hardware evolve.
Finally, all experiments are conducted using PennyLane, a popular QML platform. Using different platforms or programming languages may lead to different results.
}

\textit{Internal validity.}
\black{
Adversarial examples are generated using NES-based black-box gradient estimation due to the difficulty of obtaining exact gradients in realistic quantum settings. Alternative gradient estimators may lead to different perturbations.
In addition, hyperparameters are selected empirically, and their effects are discussed in the ablation study.
}

\textit{Construct validity.}
\black{
When constructing different test suites, the notion of strong and weak groups is adapted from prior deep learning testing practices, while additional suite settings could be explored in future evaluations.
Furthermore, to efficiently collect fault-inducing inputs, adversarial examples are used as a proxy. However, such inputs may not fully reflect real-world faults in deep learning systems~\cite{humbatova2020taxonomy}, and they depend on the specific attack algorithms and configurations.
}

\textit{Conclusion validity.}
\black{
The randomness involved in sampling inputs and injecting noise may introduce fluctuations, which are alleviated by repeated experiments. In addition, randomness during model training can lead to performance variations. We fixed random seeds to ensure reproducibility.
}

\section{Conclusion}

\black{To address the limitations of existing testing techniques for QNNs, this paper proposed multi-granularity coverage criteria to quantify test adequacy in a practical quantum setting. The criteria are black-box and superposition-targeted, making them feasible under realistic quantum execution. We conducted a comprehensive empirical study on benchmark datasets and representative QNN models to evaluate their effectiveness, robustness, and scalability under various influencing factors. The results demonstrate both the utility and the limitations of the proposed criteria. Moreover, coverage patterns can partially reflect aspects of model expressivity and robustness. Finally, we outline several challenges and recommendations for future QNN testing, calling for more fine-grained coverage criteria and more rigorous evaluation protocols.}


\bibliographystyle{IEEEtran}
\bibliography{IEEEabrv}

\end{document}